\documentclass[aoas]{imsart}

\RequirePackage{amsthm,amsmath,amsfonts,amssymb}
\RequirePackage[authoryear]{natbib}
\usepackage{float,wrapfig,graphicx,url,xcolor}

\startlocaldefs

\def\E{\text{E}}

\def\after{\text{after}}
\def\before{\text{before}}

\endlocaldefs

\begin{document}

\begin{frontmatter}
\title{Bayesian hidden Markov models for latent variable labeling assignments in conflict research: application to the role ceasefires play in conflict dynamics}
\runtitle{HMMs for latent variable labeling assignments in conflict research}

\begin{aug}
\author[A,E]{\fnms{Jonathan P} \snm{Williams}\ead[label=e1]{jwilli27@ncsu.edu}},
\author[B,C,E]{\fnms{Gudmund H} \snm{Hermansen}},
\author[B,C,E]{\fnms{H\r{a}vard} \snm{Strand}},
\author[D]{\fnms{Govinda} \snm{Clayton}},
\and
\author[C]{\fnms{H\r{a}vard Mokleiv} \snm{Nyg\r{a}rd}}

\address[A]{North Carolina State University, \printead{e1}}
\address[B]{University of Oslo}
\address[C]{Peace Research Institute Oslo (PRIO)}
\address[D]{ETH Zurich \& Centre for Humanitarian Dialogue}
\address[E]{Centre for Advanced Study, Norwegian Academy of Science and Letters}
\end{aug}

\begin{abstract}
A crucial challenge for solving problems in conflict research is in leveraging the semi-supervised nature of the data that arise.  Observed response data such as counts of battle deaths over time indicate latent processes of interest such as intensity and duration of conflicts, but defining and labeling instances of these unobserved processes requires nuance and imprecision.  The availability of such labels, however, would make it possible to study the effect of intervention-related predictors --- such as ceasefires --- directly on conflict dynamics (e.g., latent intensity) rather than through an intermediate proxy like observed counts of battle deaths.  Motivated by this problem and the new availability of the ETH-PRIO Civil Conflict Ceasefires data set, we propose a Bayesian autoregressive (AR) hidden Markov model (HMM) framework as a sufficiently flexible machine learning approach for semi-supervised regime labeling with uncertainty quantification.  We motivate our approach by illustrating the way it can be used to study the role that ceasefires play in shaping conflict dynamics.  This ceasefires data set is the first systematic and globally comprehensive data on ceasefires, and our work is the first to analyze this new data and to explore the effect of ceasefires on conflict dynamics in a comprehensive and cross-country manner.
\end{abstract}

\begin{keyword}
\kwd{state space model}
\kwd{multistate model}
\kwd{discrete-time Markov process}
\kwd{discrete-valued time series}
\kwd{count-valued time series}
\end{keyword}

\end{frontmatter}

\section{Introduction}

Within the conflict research community HMMs have been studied from a variety of different perspectives, and with varying degrees of sophistication.  Early applications of HMMs in the literature are investigated as case studies for various countries.  They were largely motivated by a perceived need for the conflict research community to explore its data beyond what can be provided by linear models, arguing that the dynamics exhibited by these data are complex, non-linear political systems \citep{Petroff:2013,Schrodt:1997a,Schrodt:1997b,Schrodt:2006}.  A few years after the preliminary investigations of HMMs in the conflict research literature, the book chapter \cite{Petroff:2013} summarized the best practices with particular emphasis on forecasts/predictions of violence.  

Overall, the ideas about, and implementations of, HMM strategies for explaining/predicting conflict data are outdated and have not advanced beyond the ideas and strategies prescribed in the classical tutorial paper, \cite{Rabiner:1989}.  See \cite{Anders:2020JPR}, \cite{besley2021big}, and \cite{randahl2022predicting} for more recent accounts.  Typically, there has been a proposed set of theoretically motivated (unobserved) conflict states (in the range of 3-6) linked by way of an HMM for pre-processing and organizing sequences of event-coded symbols from a large repository of international news summaries (provided by an agency such as Reuters).  Locally maximum likelihood estimates are obtained from a Baum-Welch algorithm, the most plausible (unobserved) states-sequences are inferred via the Viterbi algorithm, and the fitted HMMs have been regarded as predominantly uninterpretable but useful for forecasts/predictions of future data.

Furthermore, that it is not clear whether data has been properly discretized in the existing studies.  For example, \cite{Petroff:2013} describes the ability to tailor the length/number of time intervals to the precision of the data available, and discourages aggregation of data (i.e., hours, days, weeks, etc.).  While it is true that a discrete-time Markov process can be defined on any time grid, the a priori chosen grid must apply to all observed and future data sequences.  The use of time stamps of observed data sequences, as they were actually recorded in time, requires adherence to a continuous-time Markov process.  Such a process can be modeled with a continuous-time HMM, and has been studied extensively in the disease progression literature; e.g., \cite{Satten:1996} and \cite{Williams:2020}.

Our contributions are the following.  We propose a discrete-time Bayesian HMM to make inferences on how violence dynamics evolve in time over a latent, discrete, conflict-inferred state space, motivated by the new ETH-PRIO Civil Conflict Ceasefires data set \citep{clayton_introducing_2021}, combined with the violence data from the Uppsala Conflict Data Program's (UCDP) geo-referenced event data set \citep{Sundberg2013JPR}.  We use weekly battle death counts as the emitted response variable, combined with conflict-domain-theory motivated, country-specific covariates.  In particular, we demonstrate how the semi-supervised defining and labeling of conflict-inferred states is methodologically essential to developing fundamental insights into some of the most challenging contemporary questions in conflict research, such as the effect of ceasefires on conflict dynamics.  The utility of HMM frameworks for defining and assigning labels for latent variables has also been exhibited in the recent article \cite{Anders:2020JPR} to identify territorial control during a civil war, that is intrinsically difficult (if not impossible or infeasible) to manually label.  Arguably, due to a deficiency in meaningful labels, HMM-based semi-supervised data labeling strategies could pave the way for the next decade of conflict research progress.  

Additionally, we offer a variety of inferential analyses and conclusions that can be drawn from fitting conflict data within this framework, as well as graphical tools and algorithms that could be used from a policy making perspective for predicting or characterizing intensity of violence.  With sufficient data and intervention-relevant predictors it is possible to conduct analysis using the state space sampler to quantify the effect of changes in policy (e.g., how the risk or duration of a conflict would be predicted to change if interventions are implemented).

The weekly battle death count data we consider are modeled, conditional on the underlying latent state, using a negative-binomial distribution with an AR mean structure.  This is a natural choice because conflict-related death count data are time series that are commonly characterized by both over-dispersion and zero-inflation; both are common features of many of the battle death series (there are several illustrations below). For a general introduction and overview of count time series, see for example the recent review paper \cite{Davis2021}.  We implement a Markov chain Monte Carlo (MCMC) algorithm to fit the HMM, and the repeated sampling coverage of all HMM parameter estimates is evaluated via the construction of posterior credible sets.  

Ceasefires are arrangements through which conflict parties commit to stop fighting, and all ceasefires share the same immediate objective: to stop violence \citep{clayton_ceasefire_2021}. They are a common part of intra-state conflict, each year occurring in about a third of all conflicts.\footnote{For simplicity we refer to `armed conflict' as conflict or armed conflict. We follow the UCDP/PRIO Armed Conflict Database and define an armed conflict as: `a contested incompatibility that concerns government and/or territory where the use of armed force between two parties, of which at least one is the government of a state, results in at least 25 battle-related deaths in one calendar year' \citep{Gleditsch2002JPR}. In this article we only focus on internal armed conflicts.}  Between 1989 and 2020 there were at least 2202 ceasefires across 66 countries, in 109 civil conflicts \citep{clayton_introducing_2021}.  Surprisingly, despite their frequency, it remains unclear to what extent ceasefires really work, i.e., it is not known to what extent they shift a conflict from a more violent to a less violent state.    To illustrate this point, in early 2003 a ceasefire between the government of Sudan and the SPLA/M marked a transition from a long period of sustained violence into a relatively non-violent state that remained in place until the parties reached a comprehensive peace agreement in 2005. Yet in Syria, for example, where there have been more than 130 ceasefires in the ongoing civil conflict, many ceasefires seem to have produced an escalation rather than deescalation in violence, or had no effect at all \citep{karakus_between_2020}. From existing research it is not possible to determine if the Syrian or Sudanese case are indicative of the general effect of ceasefires on conflict violence.\footnote{In contrast, a burgeoning body of literature explores the drivers of ceasefire onset, and the impact that ceasefires have on other outcomes such as peace processes, crime, and state-building; see, \cite{akebo_ceasefire_2016, waterman_ceasefires_2020, bara_understanding_2021, clayton_ceasefires_2020}.}  

The lacuna in understanding the role ceasefires play in conflict is a result of conflict researchers lacking not only the necessary data but also the statistical tools to properly model violence dynamics, and to study and understand the covariates that influence these dynamics.  To date, ceasefire research is largely limited to case studies \citep{palik_watchdogs_2021, pinaud_home-grown_2020, akebo_ceasefire_2016}, or analysis tailored for the policy and practice community \citep[e.g.][]{brickhill_mediating_2018, buchanan_ceasefire_2021}. The research in this area details a number of cases in which ceasefires have ultimately proved successful \citep[e.g.,][]{de_soto_ending_1999}, but also shows that in many cases violence does not end with the onset of a ceasefire 
\citep{Kolaas:2011JPR,Jarman:2004TPV,Hoglund:2005CW,Akebo:2016}, and some ceasefires even make violence worse \citep{Luttwak:1999,Kolaas:2011JPR,Mahieu:2007IO}.  
In-depth qualitative analysis has many advantages \citep[see,][]{George2005}, but is ill-suited to systematically identifying broad trends, such as whether ceasefires in general produce a significant shift in violence dynamics. This instead requires comparative quantitative analysis which has, to date, been limited for questions surrounding ceasefire onset \citep{clayton_ceasefires_2019} and design \citep{clayton_logic_2021}. There is some evidence that ceasefires stop violence \citep{Fortna:2003IO,Fortna:2004Book}, but this is limited to inter-state conflict and the analysis suffers a number of serious methodological limitations.  Accordingly, perhaps the most fundamental question on ceasefires remains largely unanswered: Do ceasefires stop violence?

Furthermore, stopping violence can serve various purposes. Firstly, ceasefires can help to support conflict management efforts: creating breaks in the fighting to facilitate humanitarian assistance \citep{aary_concluding_1995}; helping to contain conflict when resolution is not yet possible \citep{clayton_ceasefires_2020}; and terminating violence in such a way that does not require the resolution of the incompatibility \citep{hanson_live_2020, Kreutz:2010JPR}. Second, ceasefires can also help with conflict resolution efforts: helping to build trust \citep{akebo_ceasefire_2016}; signalling control and cohesion \citep{hoglund_tactics_2011}; and creating an environment more conducive to negotiations \citep{smith_stopping_1995, Mahieu:2007IO, chounet-cambas_negotiating_2011, clayton_logic_2021}. Third, not all ceasefires are conceived for peaceful purposes. Ceasefires can be used to gain some strategic advantage, including buying time to rearm and regroup \citep{clayton_strategic_2020, smith_stopping_1995}, to support state building efforts \citep{woods_ceasefire_2011, sosnowski_ceasefires_2020}, or undertaking illicit activity \citep{Kolaas:2011JPR, dukalskis_why_2015}.\footnote{A ceasefire might prove to be successful according to one purpose (e.g., humanitarian aid), but unsuccessful in another (e.g., promoting peace talks), or successful in the eyes of one conflict party, while representing an abject failure in the eyes of another. Ceasefires might also achieve their purpose, but produce other unintended effects (e.g., promoting the splintering of a non-state group \citep[]{Plank2017}).}
Nevertheless, in almost all cases, it is logical to assume that in order to achieve their purpose, ceasefires must first achieve the immediate objective, i.e., shift conflict from a violent to a non-violent state. 

Building on existing conflict research literature, we discuss how our analyses and inferences are motivated from and translate to the existing theory for how ceasefires shape conflict violence.  Most notably, a major finding of ours is surprising evidence for an escalation in the state of violence in the pre-ceasefire period (i.e., the two weeks prior to a ceasefire) of a conflict.  

The remainder of the paper is organized as follows. In the next section, we describe the data set that motivates our work, along with a discussion of why we focus on a weekly resolution of battle deaths.  After commenting on limitations and challenges of using these data for modeling conflict dynamics, Section \ref{methods} motivates the specification of our count-valued time series model for weekly battle deaths.  The technical details of the discrete-time HMM and its implementation are provided in Section \ref{methods:hmm}.  We illustrate why simpler statistical models are inadequate for our methods with various empirical studies in Section \ref{empirical_studies}, along with a simulation study on synthetic data for the model we propose.  Results, analyses, and robustness considerations are provided in Section \ref{results}.  The paper concludes in Section \ref{conclusion} by motivating a variety of open problems in conflict research to attract the attention of other statisticians.  Documented \verb1R1 code for the use of our methods by other researchers and policy analysts on their data, along with the workflow for reproducing our results are available in the Supplementary Material \citep{williams2024} and at \verb1https://jonathanpw.github.io/research.html1.

\section{Data}\label{data}

Our goal is to build a model that is able to capture and re-create the intensity of conflict, with its spikes and lulls, as well as more enduring patterns of violence.  The violence data we utilize come from the UCDP geo-referenced event data set \citep{Sundberg2013JPR}, which reports all events with at least one battle-related casualty.  Each event record shows where and when an event took place, which actors where involved, and how many battle-related deaths ensued.  The response variable that we consider records the number of people killed due to intrastate conflict and/or internationalized intrastate conflicts per week in each country. This also means that the data for countries with several parallel on-going conflicts are collapsed into one country time series.

For ceasefires, we rely on the ETH-PRIO Civil Conflict Ceasefires data set \citep{clayton_introducing_2021}, which represents the first systematic and globally comprehensive data on ceasefires. Our work is the first to use this new data to explore the effect of ceasefires on conflict dynamics in a comprehensive and cross-country manner. The ceasefires data defines a ceasefire as `an arrangement that includes a statement by at least one conflict party to stop violence temporarily or permanently from a specific point in time'. This broad conceptualization of a ceasefire captures the full range of security arrangements through which belligerents might agree to temporarily suspend and/or terminate hostilities.
We include unilateral and bi/multi-lateral ceasefires. A unilateral ceasefire occurs if one group alone declares the cessation of hostilities. For example, in December 2018 the Tatmadaw (army) in Myanmar declared a unilateral ceasefire towards a number of armed ethnic organizations that was not reciprocated. A bi/multi-lateral ceasefire occurs when two or more parties jointly declare a ceasefire towards one another. For example, in November 2018, Israel and Hamas jointly agreed to simultaneously halt hostilities.

Focusing on the impact of ceasefires on violence dynamics, rather than peace agreements, we consider only non-definitive ceasefires (i.e., ceasefires that attempt to suspend rather than permanently terminate a conflict).  Moreover, since we are focused on country level dynamics, we also exclude ceasefires that only cover a part of the conflict area (i.e., so-called local ceasefires), as these agreements seek to reshape violence in a limited area, and so it does not make sense in our context to assess their impact on the conflict as a whole. Finally, we exclude ceasefires that extend or renew prior agreements, based on the assumption that any shift in violence dynamics is likely to have occurred in response to the original agreement.  

The ceasefires data includes the date on which the arrangement enters into effect. We define the week containing the start date of the ceasefire, together with the following four weeks (i.e., five weeks in total) as a \emph{ceasefire period}. We do this to focus the analysis and the attention of the model on the dynamics that we are most confident relate to the ceasefire. This helps to mitigate the record-keeping uncertainty surrounding the effective duration of a ceasefire. Further, we define the two weeks prior to the start of a ceasefire (i.e.\ the two weeks before the week that contains the start date) as the \emph{pre-ceasefire period}. Ceasefires tend to be negotiated fairly quickly, and once agreed to often take a few days to implement. Thus, we believe two-weeks represents a sufficiently long period so as to capture the direct period in which the ceasefire is under consideration, but short enough to avoid picking up other conflict-related factors.\footnote{Since we are aggregating the data at a country level, it means that it is possible for a country to simultaneously be in a pre-ceasefire and a ceasefire period. This is relatively rare (only 244 out of a total 3872 weeks have multiple events) thus we leave modeling this challenge to future work (see Section \ref{conclusion} for additional discussion).}

There are several countries in the data set that do not contain any battle deaths or ceasefires. Out of the 170 countries, there are 74 without any battle deaths, and 124 countries with less than 1000 battle deaths throughout the entire time period 1989--2018. These countries are important for estimating the baseline state of `non-violent', also referred to as `state 1'. We partially label the data using the following definition: a week is labelled as state 1, without error, if it is at least 60 days (in both directions in time) removed from an observation of at least one battle death, {\em and} if it is also part of a consecutive period of at least 2 years without any battle-related deaths. This is the only state labeling assignment used for model estimation.

\subsection{Control covariates} \label{control_variables}

There are a collection of standard covariate types that are known in the conflict research community to affect the likelihood of conflict, such as political regime, economic development, and population.  For these covariate types we use the following.  Political regime is measured with the polyarchy index from the V-Dem project \citep{coppedge_v-dem_2019}, used to control for alternative conflict management opportunities within Dahl's \citet{Dahl:1971} conceptualization of democracy. It is understood that countries somewhere in the middle of the polyarchy spectrum are most vulnerable to violent conflict, and so we include polyarchy in nominal value, as well its squared and cubed values, as covariates in our analysis.  The V-Dem indices consist of a mix of variables measured either at the end of a year, at the maximum value throughout a year, or as the average over a year. That being true, we use lagged polyarchy values to prevent mixing cause and effect.

Economic development is measured as gross domestic product (GDP) per capita.  GDP relates directly to the feasibility of armed conflict because potential rebels in wealthy countries have more to lose from a rebellion, and wealthy governments have a larger capacity to co-opt a population through public goods and coerce a potential rebel through a strong security apparatus. Poor countries are more likely to have poor citizens, often more willing to engage in risky warfare, less ability to co-opt, and weaker militaries.

Population affects both the risk of conflict in the first place and the likelihood of a ceasefire in a specific conflict. Larger countries are more likely to have conflicts simply because of the larger number of people able to start a rebellion. Consequently, larger countries are more likely to have several parallel conflicts, which creates a complex situation where a rebel group might seek a ceasefire with the government to actively harm competing rebel groups. The economic development and population variables are obtained from the World Development Indicators (\url{http://data.worldbank.org/indicator}), and are included in log-lag-values in our analyses.  Lastly, we include as covariates, indicators for ceasefire and pre-ceasefire periods (as defined in Section \ref{data}) for each country/week.

\subsection{Limitations of the data and weekly resolution}

The UCDP geo-referenced event data set has a number of known limitations \citep[see also][]{raleigh2023political}. The main problem is missing data, as some events are undetected and some not found newsworthy or politically useful \citep{Dawkins2021}.  Significant biases arise in UCDP data sets because they are based on secondary sources.  \cite{Weidmann2015} is the best reference for this; the bias is substantial and correlated with cell phone coverage.  Sources are sometimes uncertain, conflicting, or partially overlapping, and measurement error can occur on a number of dimensions.  An ``event size bias'' as investigated in \cite{price2014} is also an important artifact that arises when the probability of an event being reported is associated with the size of the event.  

Further, reports may be conflicting about the severity of a given event; reports may be partially overlapping, raising questions about whether there were actually two different events or imprecise reports of a single event.  Methods and case studies for determining the unique reported events (i.e., ``unique entity estimation'') and linking duplicate reported events (i.e., ``duplicate detection'' or ``entity resolution'') have been published in the statistics literature \citep[e.g.,][]{sadinle2014,sadinle2018,chen2018}.  All three papers, \cite{sadinle2014}, \cite{sadinle2018}, and \cite{chen2018}, are indeed very interesting and address a major problem in the collection of conflict data, namely the potential overlap between reports from different sources. See \cite{Brunborg2003} for a more in depth discussion of this problem in the context of Srebrenica and the requirements for legal documentation used in an international court of law. Yet the paper only addresses one dimension as it focuses on individual, civilian casualties collected by individual traits. The vast majority of casualties in conflict remain anonymous and are referred to in relation to the location and organizations involved, which introduces two dimensions not discussed by any of the papers \cite{Brunborg2003}, \cite{sadinle2014}, \cite{sadinle2018}, or \cite{chen2018}. 

Despite these limitations, UCDP has decades of experience in consistently coding such overlaps and has fairly good routines to handle data reporting challenges.  As a result, their data set contains, in addition to the best estimate, a high and low estimate.  In about 80\% of the cases these are the same or very close, but sometimes there is a substantial variation.  While this does not address missing data, it does offer an opportunity to assess the robustness of empirical findings.  Moreover, it is reasonable to believe that larger events are more likely to be precisely reported than smaller events \citep{price2014}, and that more violent periods in general are more likely to be reported.  These presumptions suffice for reasoning that the UCDP data reflect, to some extent, true {\em trends} in conflict intensity, and our studies support our assumption that overall trends in conflict dynamics are consistently measured by UCDP.  The same logic is used in \citep{Tai2022} to support similar use of event data.  Additional issues arise from conflicting reports about organizational affiliation, which is a major problem in many situations but not ours.  Since we aggregate fatalities at the country level, the organizational aspect is not relevant \citep{lacina2012}.

Aggregation to a weekly resolution of data is commonly used for the study of conflict dynamics \citep{wood_2014, Krtsch_2021, Holterman_2021}, as well as for studying other aspects of contentious behaviors such as electoral violence \citep{Reeder_2018} and denial of service attacks \citep{Lutscher_2020}.  Studying conflict data on a weekly resolution balances several concerns.  The UCDP geo-referenced event data set is often coded on a daily precision level, but a significant number of observations have some level of imprecision.  A country-day data structure would saturate the data set with large number of artificial-zero observations, which would add more noise than information.  On the other hand, a monthly aggregation would obscure much of the dynamics we aim to capture.  In our aim to trace an effect of ceasefires on conflict dynamics, the weekly resolution provides enough precision and enough time between observations.  The ceasefire data set is also reliably available at a weekly level of resolution.

There sometimes exist conflicting reports on the timing of a given event, but again, UCDP will inform about the uncertainty of temporal data.  Since a week is somewhat arbitrary one could imagine that our analysis would look slightly different if we defined a week as Thursday--Wednesday rather than Monday--Sunday, but based on a rudimentary analysis, we do not believe this particular problem to be very significant.  In the UCDP geo-referenced event data, the event start and end dates cross over into the next week in less than 10\% of the cases.

Finally, there is a limitation from only relying on incidence of battle deaths as a response variable associated with underlying conflict dynamics.  Conflict dynamics surely exhibit multifaceted manifestations, including but not limited to deaths, e.g., injury and disability.  In fact, there has been much debate on the topic of conflict related injuries; it is argued in \cite{fazal2014} that a decline in war casualties has been due to medical improvements rather than a decline in the number of wars, but adequate data does not exist to fully investigate the issue.  A paper published in 2003 --- \cite{ghobarah2003} --- did report long-term effects from civil wars on a measurement called {\em disability adjusted life years (DALY)}, but DALY is a theoretically constructed index that is not regularly available (i.e., weekly, monthly, etc.).  We do not consider counts of injuries and disabilities due to conflict simply because of a lack of adequate data on such incidences.

We admit that there are significant problems with the UCDP data, and UCDP uncertainty is sometimes, perhaps even often, present on several dimensions at once.  They have sometimes resolved this by clustering a number of unclear events together to a super-event that may last for quite some time, raising obvious issues for our weekly model.  We believe, however, that we use the best data available, and that the conceptual validity of focusing on fatalities as an indicator of conflict intensity is defensible as the best option available.  Moreover, we trust that the experts at UCDP are handling these issues in a manner which we cannot improve on.

\subsection{Challenges of modeling conflict dynamics}

\begin{figure}[H]
\centering\singlespacing
\includegraphics[scale=.4]{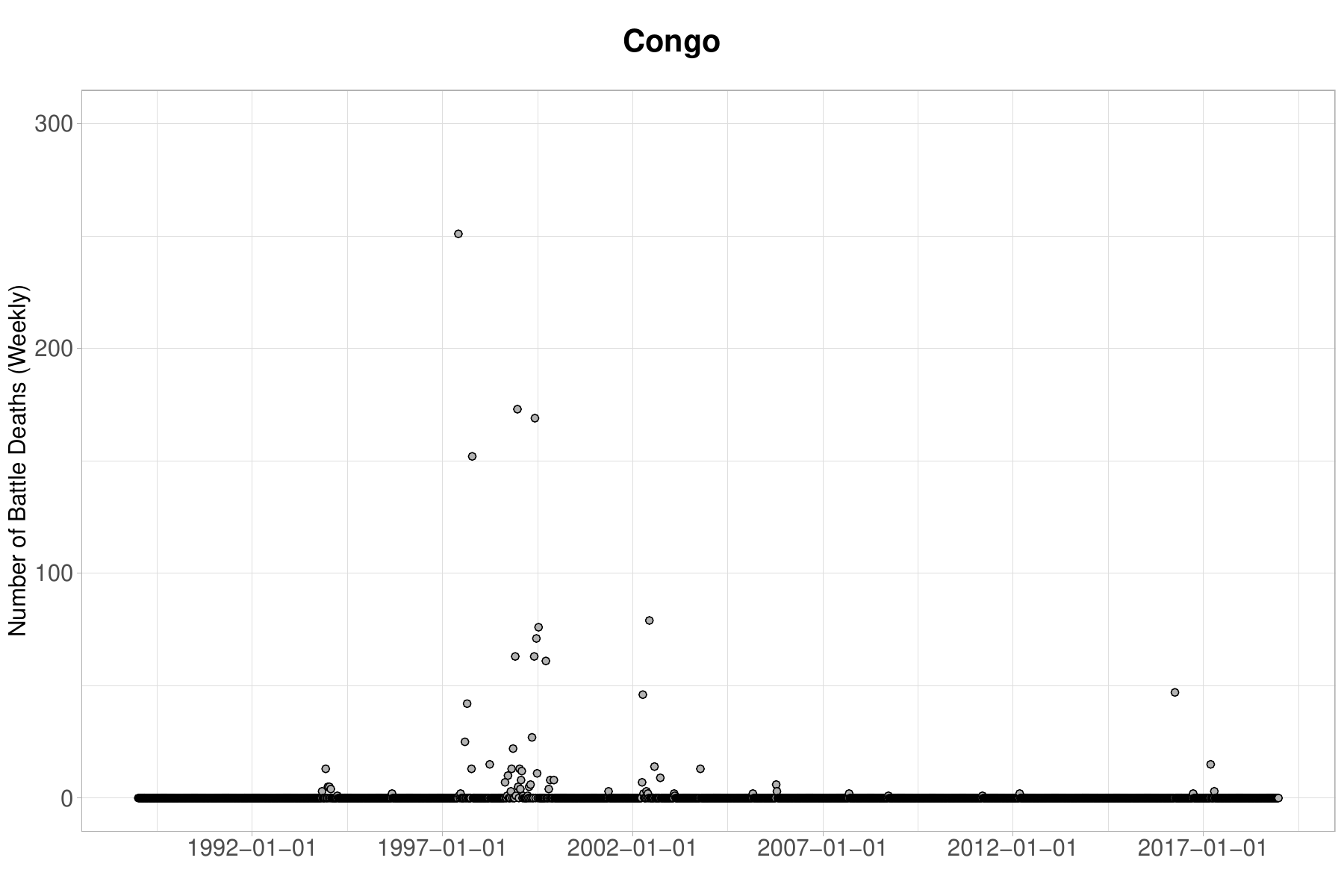}
\includegraphics[scale=.4]{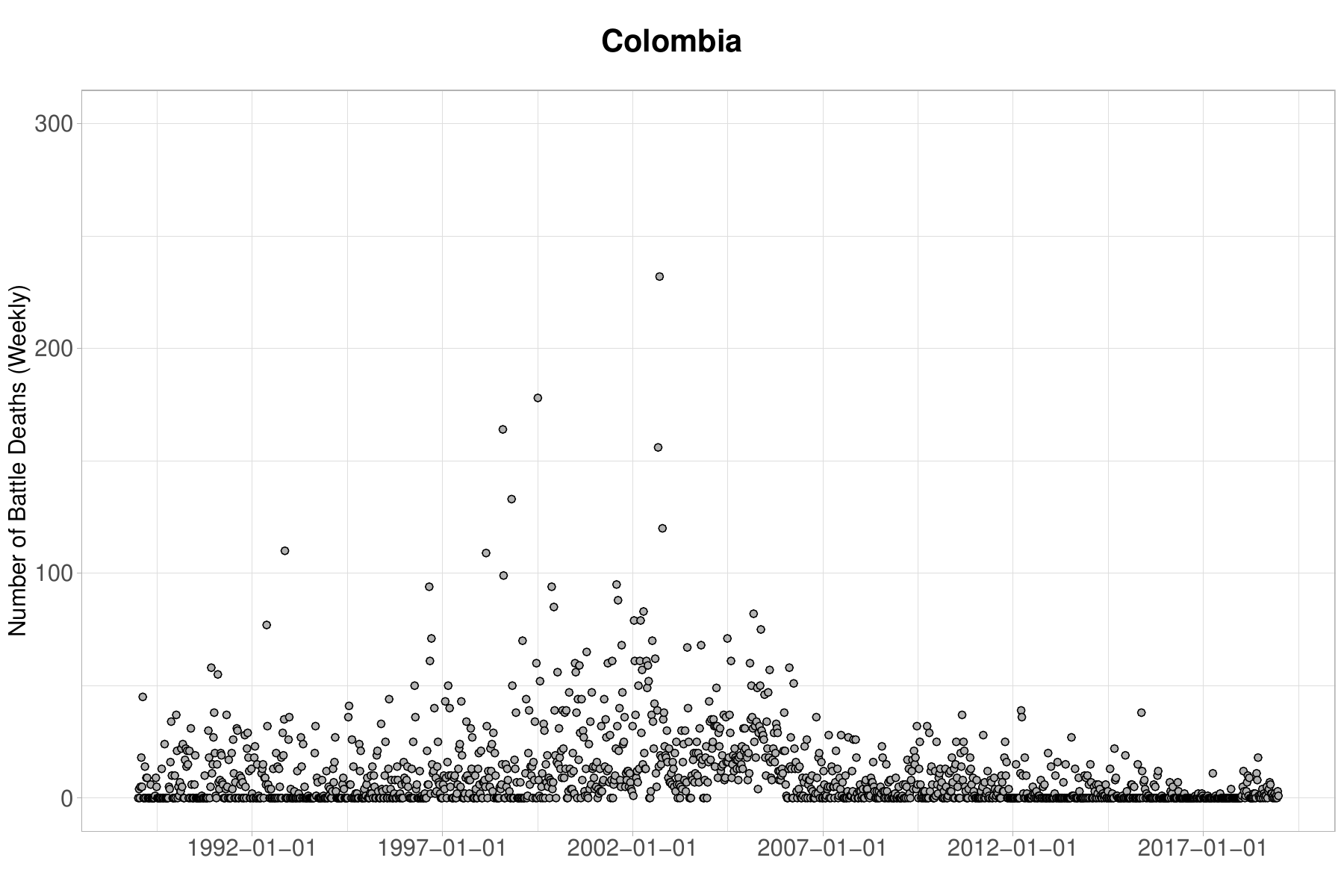}
\caption{\footnotesize Weekly number of battle deaths in the Democratic Republic of Congo ``Congo'' and Colombia, 1989--2018. Note that the Democratic Republic of Congo suffered 3,000 deaths the week of December 14, 1998, a value beyond the plot range chosen for illustration.} \label{fig:UCDPdaily}
\end{figure}

Figure \ref{fig:UCDPdaily} illustrates the statistical challenge at hand. It shows weekly aggregated number of battle deaths in internal armed conflicts over the 1989 to 2018 period for the Democratic Republic of Congo ``Congo'' and Colombia.  Once again, our goal is to build a model that is able to capture and re-create the intensity of conflict, with its spikes and lulls, as well as more enduring patterns of violence. The typical statistical models employed by conflict researchers, overwhelmingly logistic or ordinary least squares regression, count models such as Poisson and negative-binomial, and time-to-event models such as Cox proportional hazard models, are not particularly well suited for such purposes \citep[a recent review is][]{Davenport:2019ARPS}.  In contrast to the literature on conflict dynamics, the literature on the onset of armed conflict has benefited tremendously from having a, more or less, standard statistical model (a logistic cross-section time series model) and a standard set of covariates (in particular related to socio-economic development, political institutions, and demography). This standard model has allowed the literature to accumulate knowledge as more and more pieces of the puzzle have been added. Unfortunately, this has not been the case for the conflict dynamics literature, which has developed in a much less coordinated fashion.

\section{Statistical methodology}\label{methods}

\subsection{Response variable model}\label{methods:response}

Fitting a Poisson model is perhaps the simplest approach for inference on count-valued data, such as the number of battle deaths in a week, for a given country.  A suggested by Figure \ref{fig:UCDPdaily}, however, weekly battle deaths in a given country exhibit obvious patterns of zero-inflation and over-dispersion, and so a more flexible (i.e., two parameter) negative-binomial model is a standard, pragmatic choice.  Furthermore, beyond issues relating to zero-inflation and over-dispersion, temporal correlations in the week-to-week fatality counts cannot be ignored.  For instance, in \cite{jakobsen2021}, it is observed that several countries have battle death time series with fairly strong autocorrelation, motivating the application of a count time series model.

An introduction to count time series models can be found in \cite{Weiss2018}, and the recent review paper \cite{Davis2021} is a more general review.  In short, count time series models are a rich class of models, well-suited for data that are zero-inflated, over-dispersed, and with heteroscedasticity.  A variety of such models are explored in \cite{jakobsen2021} in the context of battle deaths time series. From this, there is not a clear ``best'' model for such time series; different types of models work well, provided they have sufficient capacity to model autocorrelation, zero-inflation, and over-dispersion.  Indeed, a Poisson-based model is found to be lacking a sufficient degree of dispersion, while a negative-binomial is recognized as more appropriate. Moreover, it reasoned that an AR type of (mean) structure is sufficient for capturing the relevant autocorrelation across time. 

Accordingly, to model the battle death counts $Y_{i, k}$ for week $k$ of country $i$, we specify a simple negative-binomial model with an AR mean structure:
\begin{equation} \label{nb_model}
Y_{i, k} \mid \overline{Y}_{i,(k-4):(k-1)} \sim \textrm{negative-binomial}(r_{i, k}, p),
\end{equation}
with $p = c(c + 1)^{-1}$ and 
$
r_{i, k} := a + \rho \cdot \overline{Y}_{i,(k-4):(k-1)},
$
where $\overline{Y}_{i, (k-4):(k-1)} := \frac{1}{4} \sum_{l = 1}^{4} Y_{i, k - l}$ and $c, a, \rho > 0$. 
This is actually a special case of a so-called ``NB-DINARCH(4)'' model, i.e., negative-binomial dispersion integer AR conditional heteroscedasticity model, introduced in \cite{xu2012}.  In experimenting, we determined that specifying the rate $r_{i, k}$ as an AR function of the average battle death counts --- over the previous four weeks --- balances the noise in the weekly data values with their monthly composite.  Further, this AR structure allows for meaningful interpretability of parameters with respect to the underlying conflict intensity; for example,
$$
\E\{ Y_{i, k} \, | \, \overline{Y}_{i, (k-4):(k-1)} \} = \frac{a}{c} + \frac{\rho}{c} \cdot  \overline{Y}_{i, (k-4):(k-1)}, 
$$
meaning that we can view $a/c$ as a type of baseline intensity and $\rho/c$ as an escalation parameter.  The time series is weakly stationary if $0 \le \rho/c < 1$, and explosive otherwise \citep[see Theorem 4.1]{xu2012}, which gives a natural characterization of the non-escalatory or escalatory dynamics of conflict violence.  Assuming no battle deaths occurred in a given four consecutive weeks, small values of the $c$ parameter and large values of the $a$ parameter are associated with an increased likelihood of violence on the fifth week: 
$$
\Pr\{ Y_{i, k} > 0 \mid \overline{Y}_{i, (k-4):(k-1)} = 0 \} = 1 - \Big(\frac{c}{c+1}\Big)^{a}.
$$

Model (\ref{nb_model}), however, is not fully adequate for characterizing a time series of battle deaths over the entire time period of the data (1989-2018) because it assumes that the values for $c, a, \rho$ are fixed over time and across countries.  This assumption is clearly violated for countries like Congo, where, as observed in Figure \ref{fig:UCDPdaily}, the country transitions in and out of regimes of violence and peace.  In fact, several countries in the data set do not exhibit any conflict violence for all of 1989-2018.  What is necessary is for the response variable model (\ref{nb_model}) to be able to adapt to latent regime changes between violence and peace, as well as to incorporate country-specific covariates.  These necessary extensions motivate our construction of an HMM, introduced next.

\subsection{HMM likelihood function}\label{methods:hmm}

In this section we develop our discrete-time Bayesian HMM.  As described previously, we regard the data resolution on a weekly grid, but the details would be the same for any discretization over time.  Note that if the data cannot be meaningfully organized on a grid of time points (no matter how precise), then a continuous-time HMM must be developed for the transition matrix to be properly computed \citep[e.g., see][]{Williams:2020}.

For a data set consisting of $N$ countries, denote each country by an index $i \in \{1,\dots,N\}$, and let $y_{i,k}$ be the number of observed battle deaths for week $k$, for $k \in \{1,\dots,n_{i}\}$, where $n_{i}$ is the number of weeks included in the data set for country $i$.  The value of $y_{i,k}$ results from a multitude of circumstances, and we summarize these circumstances by a latent state $s_{i,k} \in \{1,2,3\}$, corresponding to the `state' of conflict dynamic at week $k$.  These three underlying conflict-related states are defined as `non-violent', `stable violence', and `intensified violence', respectively; their names/interpretations are deduced from our inferences on the HMM fitted to the real data, as discussed in Section \ref{results}.  We limit our focus to three states, as \cite{Petroff:2013} argues that beyond three states [in an HMM of conflict dynamics] it becomes exceedingly difficult to interpret the empirical results, and finds distinctions accounted for by including additional states to be vague at best.

The underlying state sequence, $s_{i,1},\dots,s_{i,n_{i}}$, defines a stochastic process, and we assume for computational feasibility, as required for an HMM, that it is a Markov chain in that the state of the process at week $k$ only depends on the state of the process (and possibly covariates) at week $k-1$.  That being so, the conditional distribution of the random variable $S_{i,k} \mid S_{i,k-1}$ is determined by a $3\times 3$ transition probability matrix $P^{(i,k)}$, which we express as,

{\singlespacing
\[
P^{(i,k)} :=
\begin{pmatrix}
\frac{1}{1+e^{q_{1}^{(i,k)}}+e^{q_{2}^{(i,k)}}} & 0 & 0 \\
0 & \frac{1}{1+e^{q_{3}^{(i,k)}}+e^{q_{4}^{(i,k)}}} & 0 \\
0 & 0 & \frac{1}{1+e^{q_{5}^{(i,k)}}+e^{q_{6}^{(i,k)}}} \\
\end{pmatrix}
\begin{pmatrix}
1 & e^{q_{1}^{(i,k)}} & e^{q_{2}^{(i,k)}} \\
e^{q_{3}^{(i,k)}} & 1 & e^{q_{4}^{(i,k)}} \\
e^{q_{5}^{(i,k)}} & e^{q_{6}^{(i,k)}} & 1 \\
\end{pmatrix},
\]
}where $q_{1}^{(i,k)}, q_{2}^{(i,k)}, q_{3}^{(i,k)}, q_{4}^{(i,k)}, q_{5}^{(i,k)}, q_{6}^{(i,k)}$ are real-valued parameters that determine the rates of their respective state transitions.  Note that the matrix on the left simply re-scales (as row-wise multivariate logistic function transformations) the matrix on the right to have unit row sums, making it a proper transition probability matrix.  The column and row indices correspond to the state space $\{1,2,3\}$.  For example, the (1,2) component of $P^{(i,k)}$ expresses the value of the probability of transition from state 1 to state 2, for any two successive weeks.  Furthermore, we express the transition probability parameters as,
\[
\begin{pmatrix}
q_{1}^{(i,k)} & q_{2}^{(i,k)} & q_{3}^{(i,k)} & q_{4}^{(i,k)} & q_{5}^{(i,k)} & q_{6}^{(i,k)} \\
\end{pmatrix} := (x^{(i)}_{k})' \zeta,
\]
where $x^{(i)}_{k}$ is a column vector of the geopolitical, region specific control covariates from Section \ref{control_variables} as well as the indicators for ceasefire and pre-ceasefire periods, for country $i$ at week $k$, and $\zeta$ is a coefficient matrix.  The feature-rich coefficient matrix $\zeta$ is a crucial component of the HMM for the purpose of studying dynamics of conflict.  In theory, the transition rate parameters can be made arbitrarily conflict specific by including as many features (i.e., covariates and coefficient parameters) as necessary.  These features determine the rate at which the underlying state of conflict evolves or ceases to progress at all.

Accordingly, for weeks $1,\dots,n_{i}$ the probability mass function of the latent state sequence $s_{i,1},\dots,s_{i,n_{i}}$ has the form,
\begin{equation}\label{state_pmf}
\ell\big(\{s_{i,1},\dots,s_{i,n_{i}}\} \mid P^{(i,k)} \big) = \pi_{s_{i,1}} \cdot \prod_{k=2}^{n_{i}}P_{s_{i,k-1},s_{i,k}}^{(i,k)},
\end{equation}
where $P_{s_{i,k-1},s_{i,k}}^{(i,k)}$ denotes row $s_{i,k-1}$ and column $s_{i,k}$ of the matrix $P^{(i,k)}$, and $\pi_{s_{i,1}}$ is the initial state probability for state $s_{i,1}$.

This Markov process defined for the latent conflict state sequences is then embedded as a structural component of the response variable model (\ref{nb_model}), such that the parameters $a$ and $\rho$ are allowed to depend on the latent-state and to be country-specific, as follows.  For week $k \in \{1,\dots,n_{i}\}$ and country index $i \in \{1,\dots,n\}$, conditional on the latent process $S_{i,k}$, the data-generating model is expressed as
\begin{equation}\label{data_model}
Y_{i,k} \mid \overline{Y}_{i,(k-4):(k-1)}, S_{i,k} \sim \text{Negative-Binomial}(r_{i,k}, p),
\end{equation}
where $p := c(1 + c)^{-1}$,
\begin{equation}\label{nb_rate}
r_{i,k} := a_{i,k} + \rho_{i,k} \cdot \overline{Y}_{i,(k-4):(k-1)},
\end{equation}
\[
a_{i,k} := a_{1}\mathbf{1}\{s_{i,k}=1\} + a_{2}\mathbf{1}\{s_{i,k}=2\} + a_{3}\mathbf{1}\{s_{i,k}=3\},
\]
and
\begin{equation}\label{rho}
\rho_{i,k} := \mathbf{1}\{s_{i,k}\ne 1\} \cdot e^{ (\beta_{1}\mathbf{1}\{s_{i,k}=2\} + \beta_{2}\mathbf{1}\{s_{i,k}=3\})'  x^{(i)}_{k} },
\end{equation}
where $a_{1}$, $a_{2}$, $a_{3}$, and $c$ are positive parameters,  $\beta_{1}$ and $\beta_{2}$ are coefficient column vectors, and $\mathbf{1}\{\cdot\}$ is the indicator function.  Denote $a := (a_{1}, a_{2}, a_{3})$ and $\beta := (\beta_{1},\beta_{2})$.  Recall that for this model to be weakly stationary, it suffices that $0 \le \rho_{i,k} < 1$.

An implication of this data-generating process specification is that if $s_{i,k} = 1$, then $Y_{i,k} \sim \text{negative-binomial}(a_{1}, p)$.  Accordingly, the parameters $a_{1}$, $a_{2}$, $a_{3}$, and $c$ describe the model for rare conflict related deaths that may occur during weeks when a country is {\em not} experiencing a substantial conflict (e.g., isolated terrorist attacks).  Moreover, this forces the rate parameter $\rho_{i,k}$ to be identified with an increased mortality rate relating specifically to a defined period of conflict (i.e., when the system is in state 2 or state 3).  With the negative-binomial rate parameter defined as in (\ref{nb_rate}),
\begin{equation}\label{exp_value}
\E\{Y_{i,k} \mid \overline{Y}_{i,(k-4):(k-1)}, S_{i,k}\} = \frac{a_{i,k}}{c} + \frac{\rho_{i,k}}{c} \cdot \overline{Y}_{i,(k-4):(k-1)}.
\end{equation}
From the definition of $\rho_{i,k}$ in (\ref{rho}), this implies that the number of conflict deaths during state 1 has a mean of $a_{1}/c$, whereas in the conflict states the mean structure is AR.

To distinguish between states 2 and 3, the constraint that $\beta_{11} \le \beta_{12}$ is imposed (i.e., baseline $\rho_{i,k}$ for state 2 does not exceed that for state 3).  This constraint helps to facilitate the identification of states 2 and 3, respectively, as associated with `stable' versus `intensified' conflict violence.  Furthermore, we also impose the constraints that $a_{1} \le a_{2} \le a_{3}$ for the purpose of state space identification.

Finally, combining components (\ref{state_pmf}) and (\ref{data_model}) gives a full likelihood function for the HMM for each country $i \in \{1,\dots,N\}$.  For efficient estimation of the parameters, the likelihood can be expressed as a marginal mass function, resulting from integrating over all possible state space sequences.  That is, 
\begin{equation}\label{marginal_pmf}
\begin{split}
p(y_{i,5},\dots,y_{i,n_{i}}) & = \sum_{s_{i,1}=1}^{3}\cdots\sum_{s_{i,n_{i}}=1}^{3}\sum_{y_{i,1}\ge 0}\dots\sum_{y_{i,4}\ge 0} p(y_{i,1},\dots,y_{i,n_{i}},s_{i,1},\dots,s_{i,n_{i}}) \\
& = \sum_{s_{i,1}=1}^{3}p(s_{i,1}) \cdots \sum_{s_{i,5}=1}^{3}p(s_{i,5} \mid s_{i,4}) \cdot p(y_{i,5} \mid y_{i,1:4}, s_{i,5})\cdots \\
& \hspace{.5in} \times \sum_{s_{i,n_{i}}=1}^{3}p(s_{i,n_{i}} \mid s_{i,n_{i}-1}) \cdot p(y_{i,n_{i}} \mid y_{i,(n_{i}-4):(n_{i}-1)}, s_{i,n_{i}}) \\
& = \pi' \cdot P^{(i,2)} \cdots P^{(i,4)} \cdot P^{(i,5)} D^{(i,5)} \cdots P^{(i,n_{i})} D^{(i,n_{i})} \mathbf{1}, \\
\end{split}
\end{equation}
where $\pi$ is the common initial state probability column vector,

{\singlespacing\footnotesize
\[
D^{(i,k)} :=
\begin{pmatrix}
p(y_{i,k} \mid \overline{y}_{i,(k-4):(k-1)}, s_{i,k}=1) & & \\
& p(y_{i,k} \mid \overline{y}_{i,(k-4):(k-1)}, s_{i,k}=2) & \\
& & p(y_{i,k} \mid \overline{y}_{i,(k-4):(k-1)}, s_{i,k}=3) \\
\end{pmatrix},
\]
}and using the negative-binomial mass function,
\[
p(y_{i,k} \mid \overline{y}_{i,(k-4):(k-1)}, s_{i,k}) = \binom{r_{i,k} + y_{i,k} - 1}{y_{i,k}} p^{r_{i,k}} (1-p)^{y_{i,k}}.
\]

Note that if further state space information is available, such as partial labels, then the number of state sequences that are integrated over is reduced.  For example, if it is known that $s_{i,k} \in \{2,3\}$ for some $k \in \{1,\dots,n_{i}\}$, then $\sum_{s_{i,k}=1}^{3}\cdot$ reduces to $\sum_{s_{i,k}=2}^{3}\cdot$ within expression (\ref{marginal_pmf}).  Equivalently, the $(1,1)$ component of $D^{(i,k)}$ is set to zero.
 
Finally, the joint posterior density including the data from all countries $i \in \{1,\dots,N\}$ then has the form,
{
\begin{equation}\small\label{posterior_density}
\pi(\zeta, \beta, a, c \mid \{y_{i,k}\}) \propto \bigg[\prod_{i=1}^{N}p(y_{i,5},\dots,y_{i,n_{i}}) \bigg] \cdot \pi(\zeta, \beta, a, c) \cdot \mathbf{1}\{\beta_{11}\le\beta_{12}, a_{1}\le a_{2}\le a_{3}\},
\end{equation}
}where $\pi(\zeta, \beta, a, c)$ is a prior density.

\subsection{Remarks on implementation}\label{sec:implementation}

We implement a Metropolis-within-Gibbs MCMC algorithm to draw posterior samples from (\ref{posterior_density}).  With these posterior samples, we then estimate the conditional posterior distribution of the latent state space for a given country in our data set; we refer to this algorithm as the {\em state space sampler} algorithm.  The state space sampler algorithm is a tool for evaluating or predicting the most likely state (i.e., non-violent, stable violence, or intensified violence) at any given country/week, based on the HMM fit to the real data.  In Section \ref{results}, we present a visual representation of the posterior distribution of the latent state sequence for countries both in our training data and in our held-out test data.

\section{Empirical studies}\label{empirical_studies}

\subsection{Illustration of why simpler statistical models are inadequate}

As motivated by Sections \ref{methods:response} and \ref{methods:hmm}, the negative-binomial response model for battle death counts with an AR mean structure, conditional on a latent state space process, is essential for capturing features of these data that are critical for pursuing conflict research questions. Moreover, assuming the latent state space process is Markovian, as in the HMM framework we proposed, is the simplest and most computationally pragmatic approach.  To exemplify the importance of these methodological features, in this section we examine the limitations of simpler approaches to analyzing these data.  Such simpler approaches may lead to similar, albeit more limited, conclusions as those that can be drawn from fitting our full HMM model (e.g., as in Section \ref{results}).  Simpler statistical approaches involve numerous subjective choices that are not necessary easy to substantiate.  Many of these subjective choices become data driven in the context of our HMM approach.

\begin{figure}[H]
\centering\singlespacing
\includegraphics[trim={0 0 0 20}, clip, scale=.45]{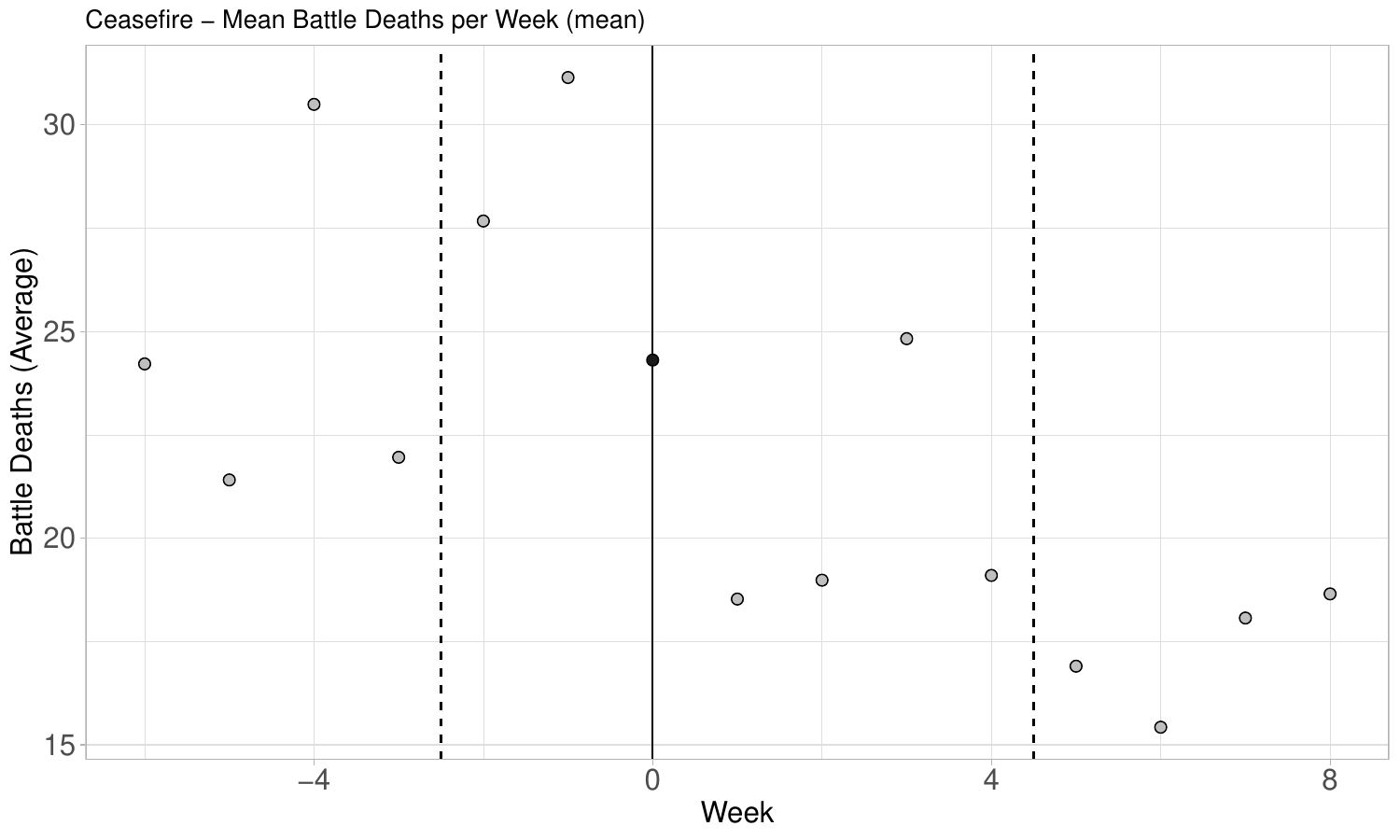}
\caption{\footnotesize Displayed is the mean number of battle deaths per week, of the 4 weeks before each ceasefire declaration and the 4 weeks after each ceasefire is in effect.  All battle deaths series are centered at the week of the ceasefire, which is then defined as week 0.  The two vertical dashed lines mark the period defined as declared and ceasefire, respectively.}
\label{fig:simple_4_week_untrimmed}
\end{figure}

\begin{figure}[H]
\centering\singlespacing
\includegraphics[trim={0 0 0 20}, clip, scale=.45]{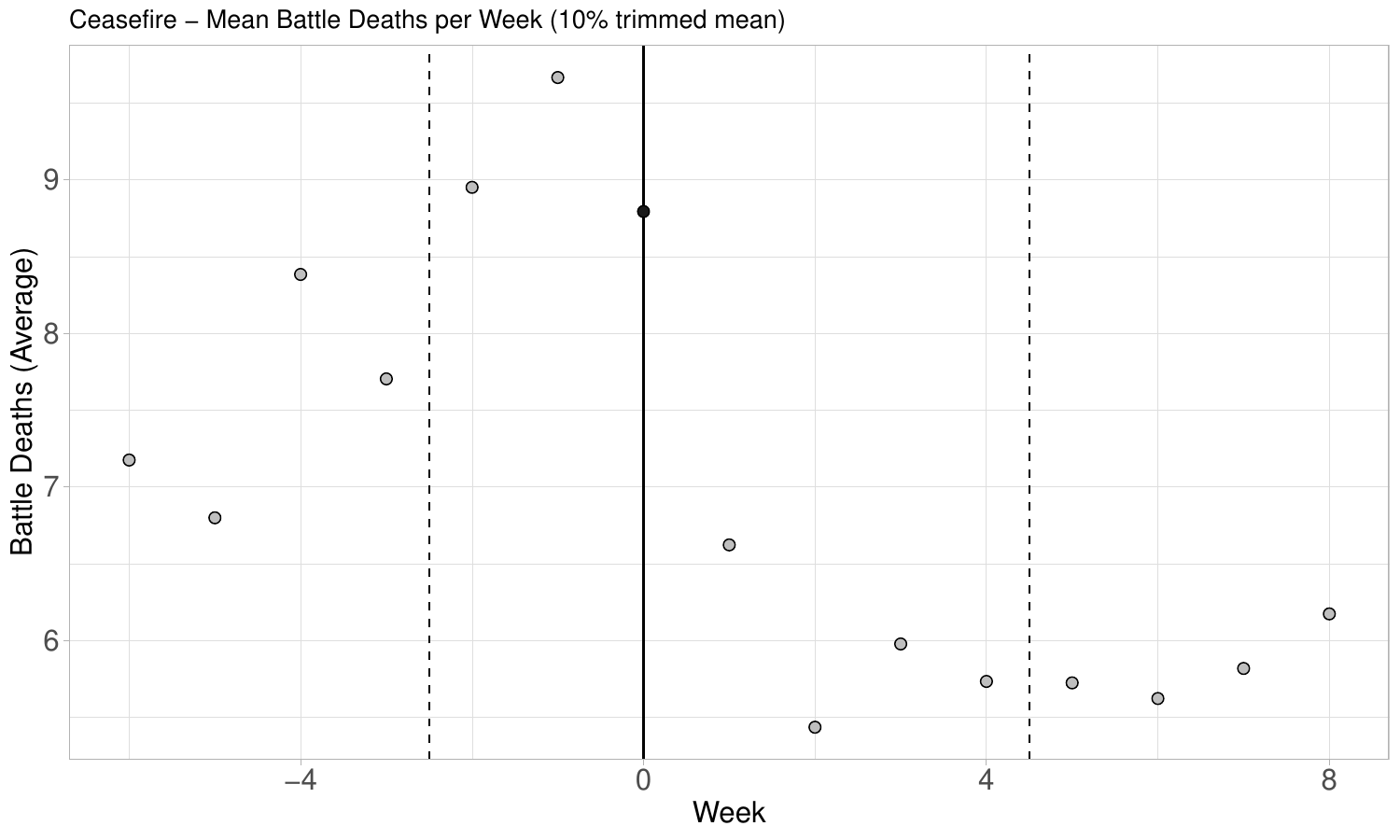}
\caption{\footnotesize Displayed is the 10\% trimmed mean number of battle deaths per week (i.e., the mean after removing the 10\% highest and lowest values for each week surrounding all ceasefires, for all countries in the data set), of the 4 weeks before each ceasefire declaration and the 4 weeks after each ceasefire is in effect.  Compare to Figure \ref{fig:simple_4_week_untrimmed}.}
\label{fig:simple_4_week}
\end{figure}

A most simple approach is to compare the number of battle deaths before and after many records of ceasefires.  If ceasefires work as intended, we would expect, on average, a statistically/practically significant reduction in the magnitude of violence after a ceasefire is in effect.  Figure \ref{fig:simple_4_week_untrimmed} plots the weekly average number of battle deaths in the 8 weeks surrounding each ceasefire recorded in the data set, over all countries; it is seen that the average number of battle deaths tend to be higher in the weeks preceding a ceasefire than those that follow.  These data, however, exhibit a right skew due to the influence of a few countries with exceptionally high battle death counts, and so we instead reconstruct Figure \ref{fig:simple_4_week_untrimmed} and use a trimmed mean for the robustness of our subsequent analyses in this illustrative section; see Figure \ref{fig:simple_4_week}.

\begin{figure}[H]
\centering\singlespacing
\includegraphics[trim={0 0 0 20}, clip, scale=.45]{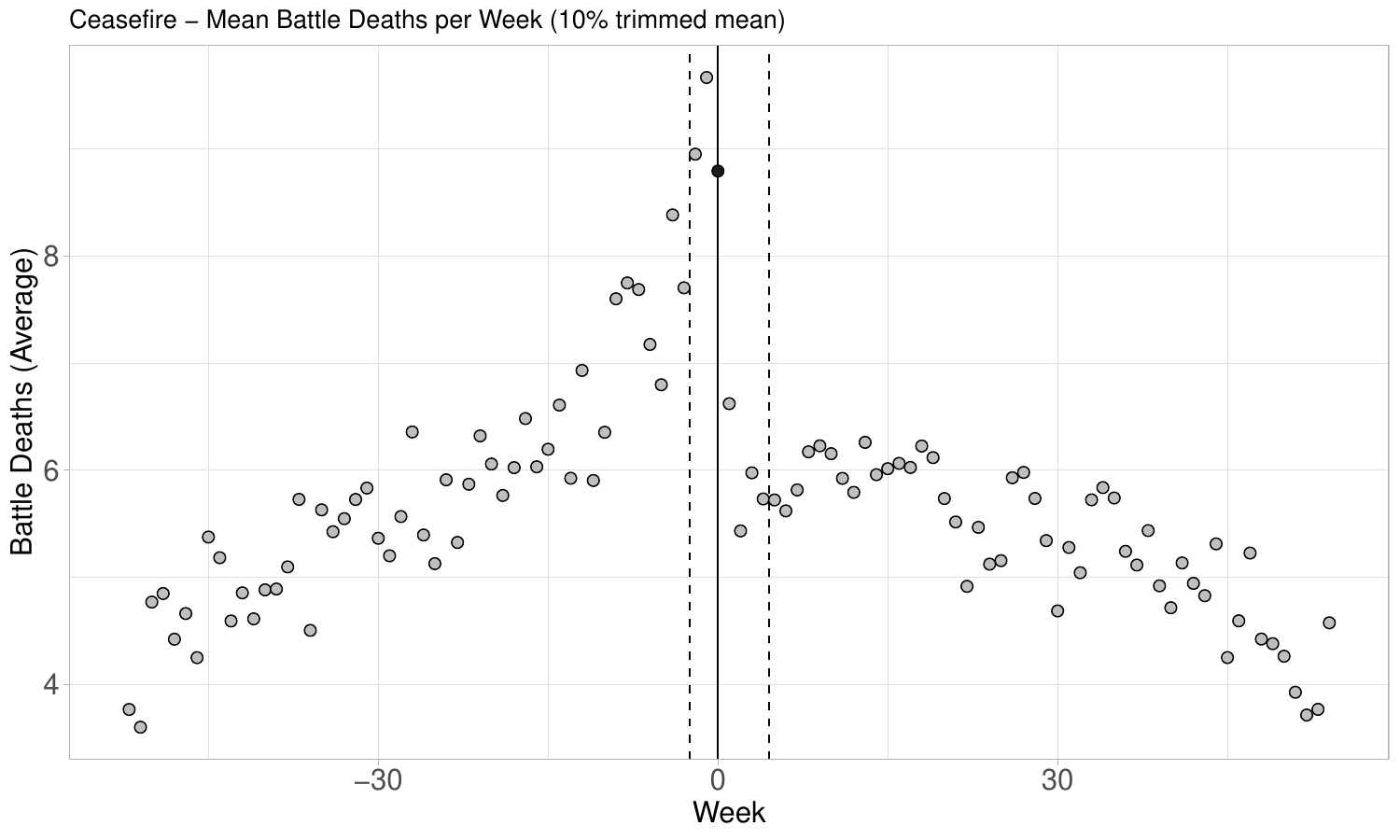}
\caption{\footnotesize The 10\% trimmed mean number of battle deaths per week, of the 50 weeks before each ceasefire declaration and the 50 weeks after each ceasefire is in effect.  Compare to Figure \ref{fig:simple_4_week}.} 
\label{fig:simple_50_week}
\end{figure}

A further complication is that the number of weeks to include in the {\em before} and {\em after} ceasefire periods are indeterminate.  Including too many weeks before or after the ceasefire will pull down the average battle death count, as illustrated by Figure \ref{fig:simple_50_week}, simply because conflicts typically expire, eventually.  The diminishing battle death counts near the right and left boundaries reflect increasingly more weeks associated with periods of peace.  Next, since there are sometimes less than, for example, 50 weeks between two ceasefires, [subjective] choices are required to avoid weeks that overlap between $\pm 50$ weeks of more than one ceasefire; e.g., keeping or removing ceasefires with overlapping intervals or reducing the number of weeks to include before and after -- all options will likely introduce a systematic bias into the analysis.  Moreover, it is unrealistic to assume a fixed number of weeks (e.g., be it 4, 50, etc.) would apply to all conflicts, across all countries and time.  The utility of the HMM framework introduced in Section \ref{methods} is that the model determines the weeks most likely associated with conflict and peace and at the country-specific resolution, so that the effects of ceasefires on violent conflict can be estimated without conflating weeks of peacetime in the estimation.  

For the sake of more precise illustration, we will proceed to compare the 10\% trimmed mean number of battle death counts before and after the ceasefires as in Figure \ref{fig:simple_4_week}, i.e., including the 4 weeks before each ceasefire declaration and the 4 weeks after each ceasefire is in effect.  Let $W_{j,k}$ be the number of fatalities associated with week $k \in \{-6, \dots, -3\}\cup\{5, \dots, 8\}$ for ceasefire $j \in \{1,\dots,J\}$, where $J$ is the total number of ceasefires observed in the data set, within and across all 170 countries.  For the simple approach of investigating the role ceasefires play in conflict dynamics by comparing battle death counts before and after a ceasefire (again for now, assuming exactly 4 weeks of counts before and after are appropriate), all that is necessary is to fit the parameters of an analysis of variance (ANOVA) model, such as: independently for $j \in \{1,\dots,J\}$ and $k \in \{-6, \dots, -3\}\cup\{5, \dots, 8\}$,
\begin{equation} \label{simple_model}
W_{j,k} = \delta \cdot \mathbf{1}\{k < -2\} + \mu_{j} + U_{j,k},
\end{equation}
where $\delta + \mu_{j}$ represents the expected number of fatalities before a ceasefire has been declared, and $U_{j,k}$ is a mean zero innovation.  In this specification, $\delta$ is the theoretical effect of the ceasefire on the expected number of battle deaths. 

Next, denote 
\begin{equation}\label{before_after}
\overline{W}_{j,\before} := \frac{1}{4}\sum_{k=-6}^{-3} W_{j,k} \quad
\textrm{ and }
\quad
\overline{W}_{j,\after} := \frac{1}{4}\sum_{k=5}^{8} W_{j,k},
\end{equation}
and
$$
\overline{V}_{j} := \overline{W}_{j,\before} - \overline{W}_{j,\after} = \delta + (\overline{U}_{j, \before} - \overline{U}_{j, \after}), 
$$
with $\overline{U}_{j,\before} $ and $\overline{U}_{j,\after}$ defined analogous to the averages in (\ref{before_after}).  In this formulation, $\E(\overline{V}_{j}) = \delta$ and the 10\% trimmed mean of the observed $\overline{v}_{j} = \overline{w}_{j,\before} - \overline{w}_{j,\after}$ for $j \in \{1,\dots,J\}$ is a seemingly robust estimate for $\delta$.  Accordingly, we find the estimated $\delta$ from the data is $\sum_{1}^{J}\overline{v}_{j}/J = 2.36$ with a corresponding 0.95 bootstrap confidence interval of (1.08, 3.79), indicating a significant decrease in fatalities after a ceasefire is in effect.  Note that removing ceasefires that have overlapping weeks in the 4 weeks before a ceasefire is declared, or the 4 weeks after a ceasefire is in effect, does not change these estimates significantly.  

We have not yet, however, controlled for any ceasefire-specific covariates, such as GDP, population, or polyarchy (democracy score).  To do so, a simple extension to model (\ref{simple_model}) is
\begin{equation} \label{simple_lin_model}
W_{j,k} = (\delta +  x_{j}'\gamma) \cdot \mathbf{1}\{k < -2\} + \mu_{j} + U_{j,k},
\end{equation}
where $x_{j}$ is a vector of ceasefire-specific covariates consisting of GDP, population, and polyarchy, and $\gamma$ is a corresponding coefficient parameter vector with 3 components.  In Table \ref{table:simple_lin_model}, we present the trimmed 10\% least squares estimates \citep[c.f.,][]{rousseeuw1984} of the parameters in model (\ref{simple_lin_model}).

\begin{table}[H] 
\centering 
\begin{tabular}{l c r r r}
\hline\hline 
 & & Estimate & 0.025 & 0.975 \\ [0.5ex] 
\hline 
Ceasefire & $\delta$ & 2.26 & 1.09 & 3.87 \\ 
GDP & $\gamma_1$ & -0.02 & -1.57 & 1.76 \\ 
population &  $\gamma_2$ &-0.21 & -2.08 & 1.21 \\
polyarchy &  $\gamma_3$ & -2.16 & -3.98 & -0.17 \\
\hline
\end{tabular}
\caption{\footnotesize Parameter estimates for model (\ref{simple_lin_model}) are based on trimmed 10\% least squares, after normalizing the explanatory variables, i.e., mean centered and scaled to have unit standard deviation.  The 0.025 and 0.975 columns display the corresponding bootstrapped quantiles.}\label{table:simple_lin_model}
\end{table}

\begin{figure}[H]
\centering\singlespacing
\includegraphics[trim={21 0 0 0}, clip, scale=.5]{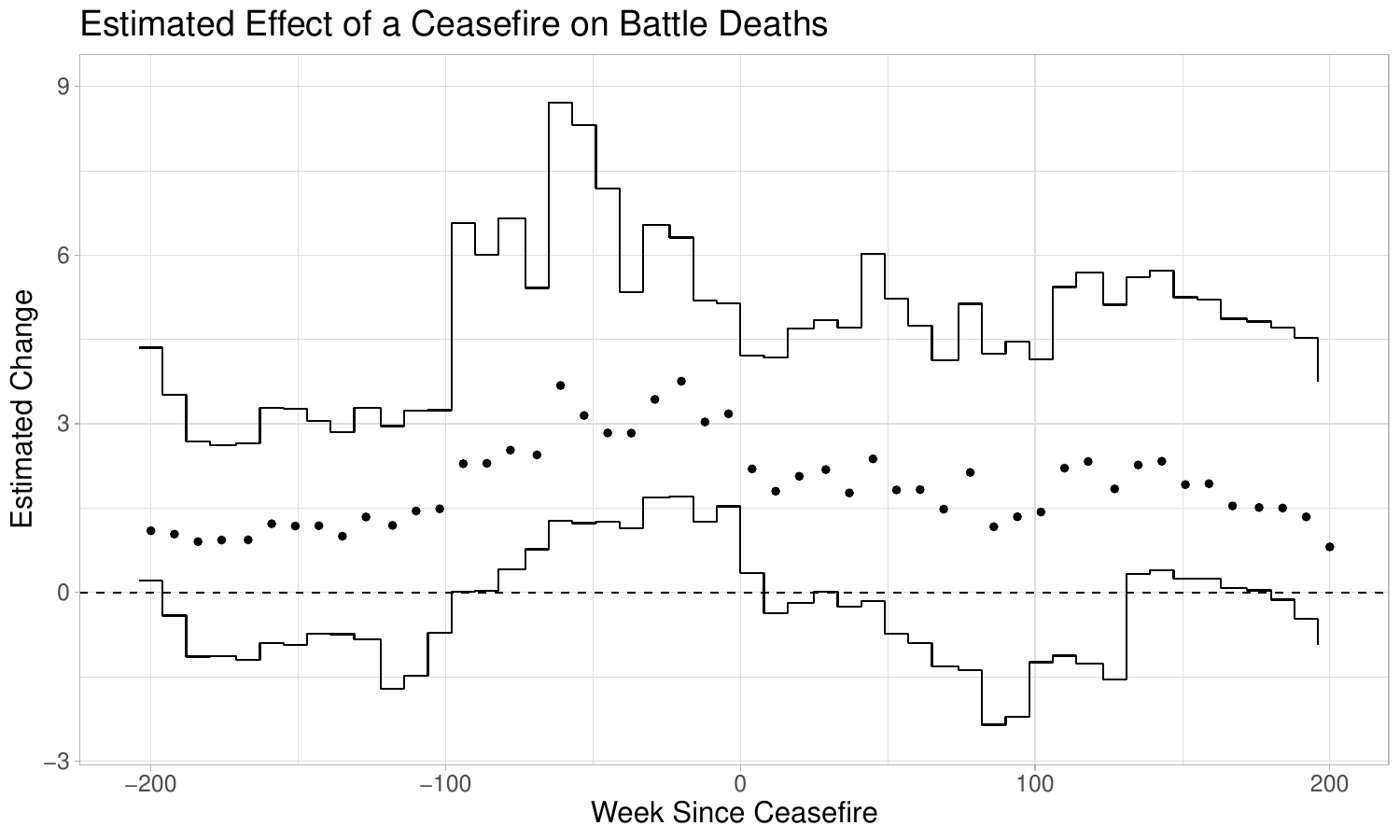}
\caption{\footnotesize The estimated effect, $\delta$ as in model (\ref{simple_model}), is plotted as points, each corresponding to a different number of weeks included in the periods before a ceasefire is declared and after it has gone into effect.  Zero ``Weeks Since Ceasefire'' corresponds to including 4 weeks before and after, as in the discussion above; the magnitude of negative (positive) values of ``Weeks Since Ceasefire'' correspond to how many additional weeks are included in the period before (after) a ceasefire.  The solid lines correspond to 0.025 and 0.975 bootstrapped quantiles.  All ceasefires that have overlapping ceasefire time periods were removed.  It is clear that the choice of the number of weeks to compare before and after a ceasefire will influence the significance of the effect of a ceasefire.}
\label{fig:simple_estimate_week_subset}
\end{figure}

\begin{figure}[H]
\centering\singlespacing
\includegraphics[trim={21 0 0 0}, clip, scale=.5]{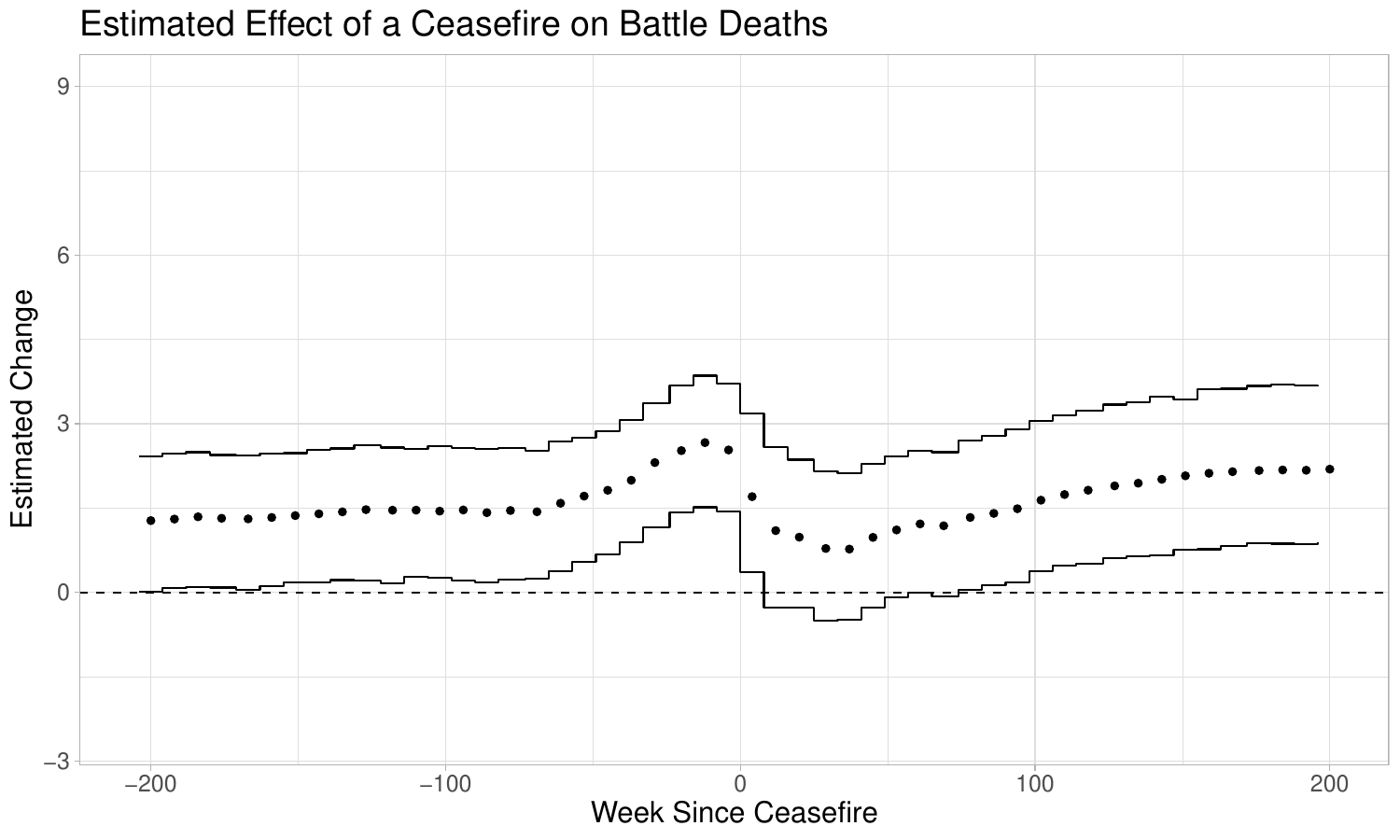}
\caption{\footnotesize Same as Figure \ref{fig:simple_estimate_week_subset}, but without removing ceasefires that have overlapping ceasefire time periods.}
\label{fig:simple_estimate_week_all}
\end{figure}

The estimates displayed in Table \ref{table:simple_lin_model} for the $\delta$ parameter are more or less unchanged from those without including covariates, and polyarchy is the only explanatory variable with evidence of a significant association, at the 0.95 level.  If we remove ceasefires that have overlapping time periods, however, then the effect of the democracy score polyarchy is no longer significant.  Moreover, Figures \ref{fig:simple_estimate_week_subset} and \ref{fig:simple_estimate_week_all} demonstrate how the estimate of $\delta$ as in model (\ref{simple_model}) will change if more or less weeks are included in the periods before a ceasefire is declared or after it has gone into effect.  These figures showcase how an investigation of the role ceasefires play in conflict dynamics depends on the number of weeks to include -- whether ceasefires with overlapping time periods are included or not; it is clear that some choices will lead to the finding of a $\delta$ significantly different from zero while other choices will not.  Beyond this finding, such an analysis does not allow for the number of before/after weeks to be ceasefire-specific.  The HMM framework we propose in Section \ref{methods} avoids these limitations because the HMM itself will quantify the uncertainty, in a data-driven manner, for the number of before/after weeks to include for each observed ceasefire in estimating the effect of a ceasefire, and at the country-specific resolution (i.e., using country-specific covariates).  Moreover, our HMM formulation estimates the effect of the presence of a ceasefire on a rolling basis, thus bypassing the need for justifications about dropping/including weeks with overlapping ceasefires that would be necessary in the simpler, ANOVA approach with predefined before/after weeks.  A problem with the ANOVA approach is that, unlike the HMM approach, it treats each ceasefire as an independent event, even if it overlaps a simultaneously occurring ceasefire.

\subsection{Simulation study of synthetic battle death data}\label{sim_study}

The purpose of this section is to verify that synthetic data generated by our HMM resembles important features of the real battle death count data, and that our Bayesian estimation procedure produces credible sets for the HMM parameters that achieve their corresponding repeated sampling coverage.  The `true' HMM parameter values used to generate the synthetic data are set as the posterior means estimated from the real data set; those values are presented in Table \ref{posterior_means}.

We generate synthetic data for each of $N = 167$ countries in our training data set (leaving data from 3 countries as test data), based on the following procedure.  For each country/week with recorded covariates a `true' latent state $s_{i,k}$ is sampled from either the initial state probability vector $\pi$ if $k=1$ or $P_{s_{i,k-1},\cdot}$ if $k>1$, and a count of battle deaths $y_{i,k}$ is sampled from (\ref{data_model}).  Note that the first four values, $y_{i,1:4}$, are sampled from (\ref{data_model}) with $\rho_{i,k} = 0$.  Furthermore, the maximum number of battle deaths in any one week is restricted to not exceed the highest week-death-count to country-population proportion of any country in the data set.  The highest such proportion is approximately 0.0006 for Congo.  If a simulated death count $y_{i,k}$ exceeds 0.0006 times the country population, then the generated data for that country is discarded and re-simulated until the restriction is satisfied.  This was not a problem for any of the 167 countries in the data set, with the notable exception of India.  For India, it took an excessive amount of time to generate battle death data that satisfy the maximum weekly death restriction, and so India was omitted from our simulation study.

\begin{table}[H]
\setlength{\tabcolsep}{1.2mm}
\begin{tabular}{p{0.1\linewidth} rrrrrrrr}
trans. rates & baseline & pre-ceasefire & ceasefire & v2x & v2x$^2$ & v2x$^3$ & GDP & pop \\
\hline
$\zeta_{1}'$ ($1\to2$) & .66 & .83 & .91 & .92 & .93 & .94 & .90 & .91 \\
$\zeta_{2}'$ ($1\to3$) & .94 & .97 & .93 & .94 & .92 & .94 & .95 & .94 \\
$\zeta_{3}'$ ($2\to1$) & .93 & .96 & .94 & .97 & .99 & .94 & .95 & .92 \\
$\zeta_{4}'$ ($2\to3$) & .95 & .96 & .97 & .93 & .91 & .97 & .95 & .93 \\
$\zeta_{5}'$ ($3\to1$) & .96 & .96 & .93 & .93 & .96 & .95 & .92 & .95 \\
$\zeta_{6}'$ ($3\to2$) & .95 & .93 & .97 & .94 & .91 & .94 & .97 & .98 \\
\\
AR coef. & baseline & pre-ceasefire & ceasefire & v2x & v2x$^2$ & v2x$^3$ & GDP & pop \\
\hline
$\beta_{1}'$ (state 2) & .90 & .97 & .95 & .96 & .95 & .96 & .92 & .94 \\
$\beta_{2}'$ (state 3) & .96 & .97 & .97 & .96 & .95 & .93 & .97 & .93 \\
\\
other & $a_{1}$ & $a_{2}$ & $a_{3}$ & $c$ & $\pi_{2}$ & $\pi_{3}$ \\
\hline
& .91 & .66 & .95 & .96 & .90 & .95  \\
\end{tabular}
\caption{Proportion of 100, .95 posterior credible sets that contains the true parameter value for each of the HMM parameters (constructed by excluding the upper and lower .025 tails of each marginal posterior distribution). Synthetic data for $N = 166$ countries are generated for each of the 100 data sets in this simulation study.  Note that parameter values reflect covariate values for lag v2x polyarchy (linear, quadractic, and cubic), lag and log GDP per capita, and lag and log population.  These variables have all been centered and scaled to have unit variance.  See the Supplementary Material \citep{williams2024} for box plots over the 100 posterior medians for each parameter.}\label{coverage}
\end{table}

India is an outlier country in our data set in the sense that it has an uncharacteristically large population size which, based on the fitted HMM parameters (see Table \ref{posterior_means}) is associated with markedly less time spent in state 1, and battle death sequences represented with an explosive or non-stationary series (i.e., $\rho_{i,k} > 1$).  Additionally, India is often in a state of ceasefire which is associated with increased instability (i.e., states 2 and 3).  One possible explanation for why conflict dynamics are not explained so well by our fitted HMM for countries with populations as large as India is the greater possibility for numerous unrelated conflicts ongoing at any point in time.  For the average country, conflict dynamics are more likely limited to a single conflict at a time.

\begin{figure}[H]
\centering\singlespacing
\includegraphics[scale=.36]{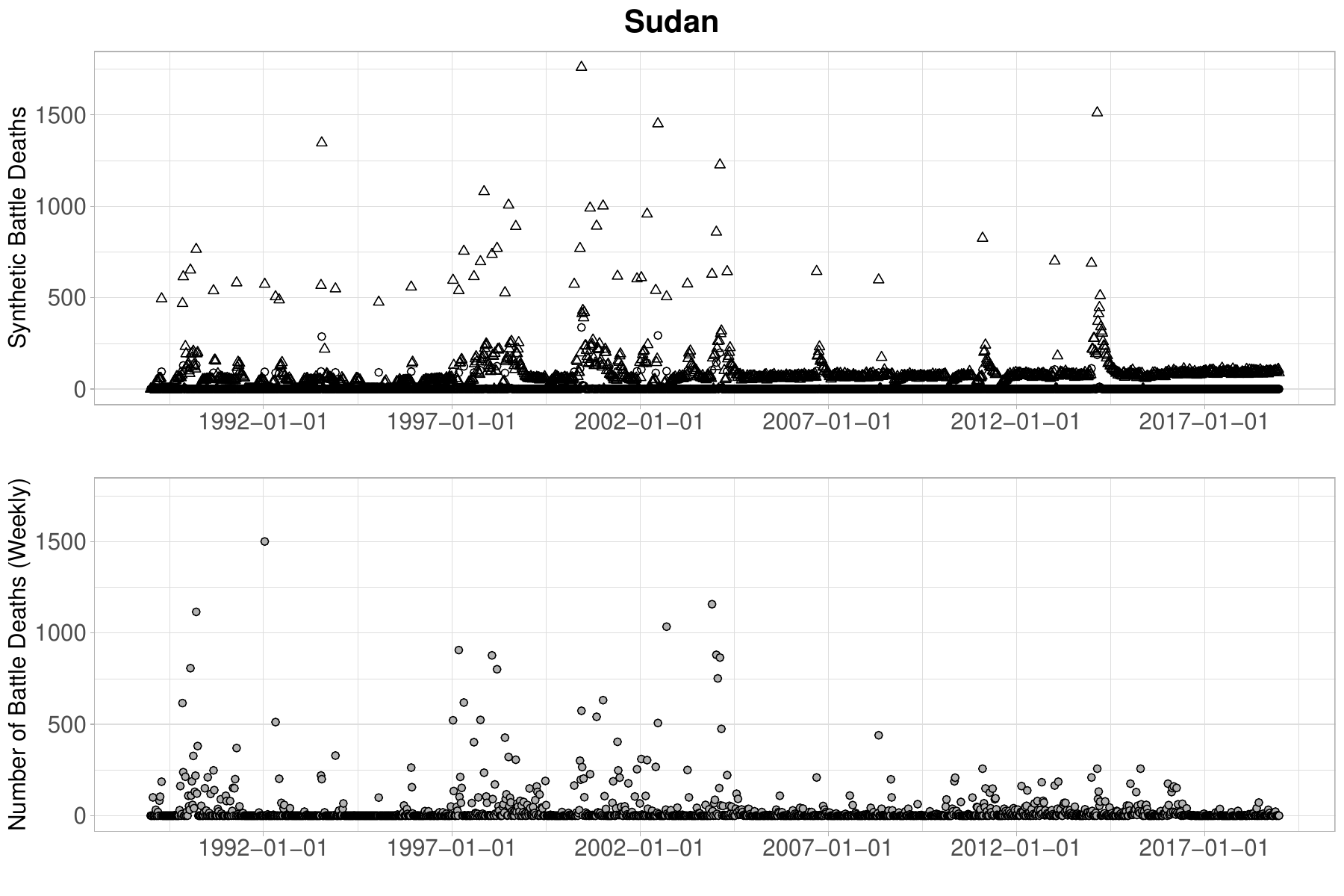}
\includegraphics[scale=.36]{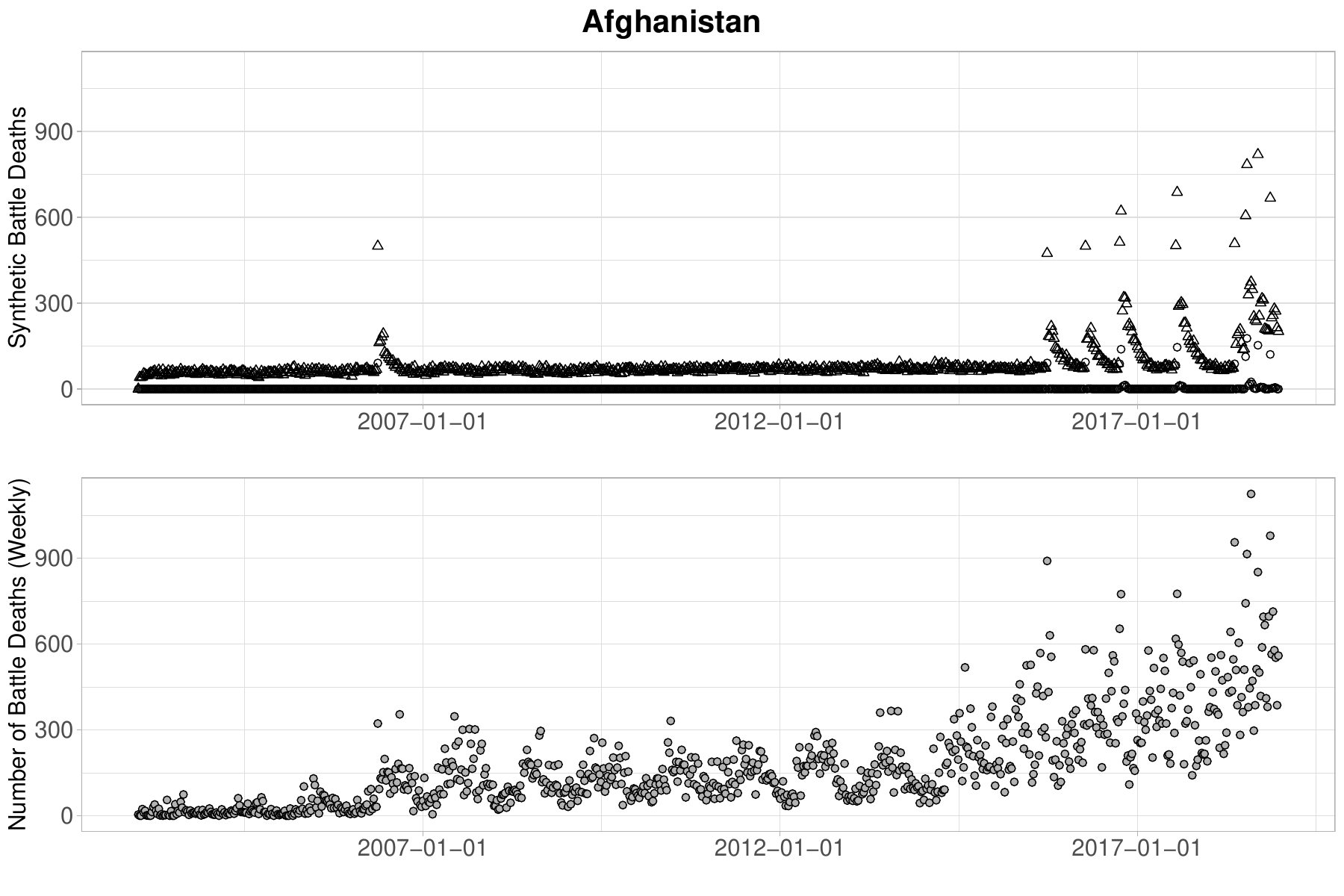}
\caption{\footnotesize The top panels for each country display 1000 synthetic realizations of the response variable sequence simulated from the HMM fit with the posterior means of all parameters presented in Table \ref{posterior_means}, the maximum a posteriori latent state sequence, and all covariate values observed in the real data set.  The triangle points represent the upper 0.025 percentile while the circle points represent the lower 0.025 percentile.  For reference, the bottom panels for each country display the real numbers of observed battle deaths.}
\label{fig:sim_test}
\end{figure}

As with our real data set, for our synthetic data sets we apply a single rule-based partial label.  That is, any week that is at least two months after or two months before a week with one or more battle deaths, and is within a two year sequence of weeks with no battle deaths is labeled as state 1, without error.  A total of 100 synthetic data sets are generated based on the described procedure.  We implement a simple independent components Gaussian prior density with mean zero and excessively diffuse standard deviation 20 for all parameters.  To enforce the constraint that $a_{1}$, $a_{2}$, $a_{3}$, and $c$ are positive-valued, we place the Gaussian prior on $\log(a_{1})$, $\log(a_{2})$, $\log(a_{3})$, and $\log(c)$.  Similarly, the Gaussian prior is placed on the logit transforms of the components of the initial state probability vector $\pi$. An MCMC algorithm is used to estimate the posterior distributions for all 70 HMM parameters, for each of the 100 synthetic data sets.  The coverage at the 0.95 level of significance for each parameter is stated in Table \ref{coverage}.  Box plots of the 100 posterior medians for each parameter are presented in our Supplementary Material \citep{williams2024}.  Figure \ref{fig:sim_test} gives a visual representation of the synthetic data we generated for the held-out test set countries, Sudan and Afghanistan.

\section{Results and analyses}\label{results}

\begin{figure}[H]
\centering\singlespacing
\includegraphics[trim={0 4in 0 0},clip,scale=.45]{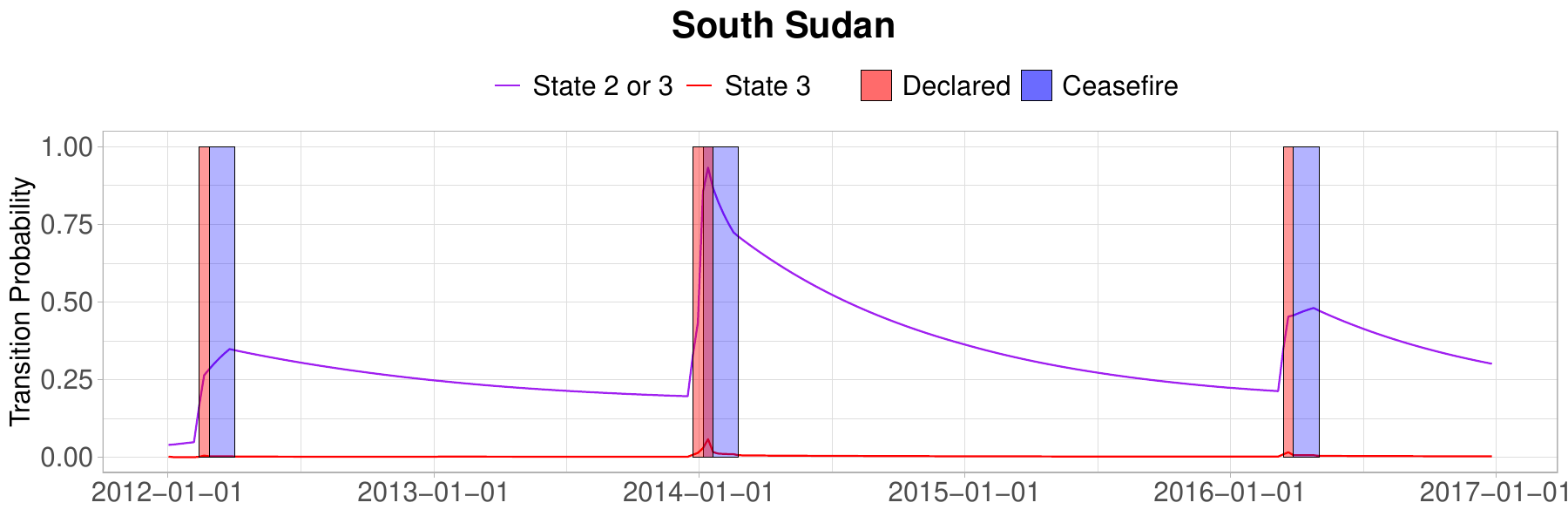}
\caption{\footnotesize Displayed is the evolution of the probabilities of transition on a week-by-week resolution.  Counts of conflict related deaths are omitted from the computation of these probabilities.  Instead, the probabilities exclusively reflect the effects, on the transition rates, of the covariates observed for the labelled country, over time.}
\label{fig:transSouthSudan}
\end{figure}

\begin{figure}[H]
\centering\singlespacing
\includegraphics[trim={0 4in 0 0},clip,scale=.45]{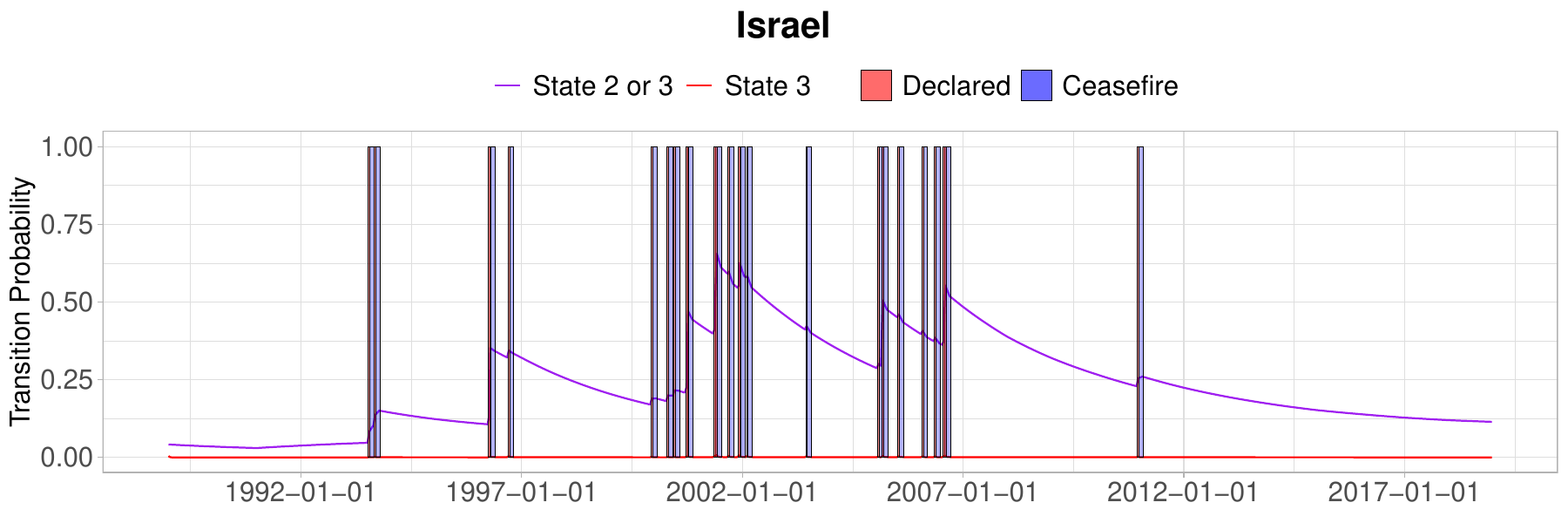}
\caption{\footnotesize Displayed is the evolution of the probabilities of transition on a week-by-week resolution.  Counts of conflict related deaths are omitted from the computation of these probabilities.  Instead, the probabilities exclusively reflect the effects, on the transition rates, of the covariates observed for the labelled country, over time.}
\label{fig:transIsrael}
\end{figure}

\begin{figure}[H]
\centering\singlespacing
\includegraphics[scale=.45]{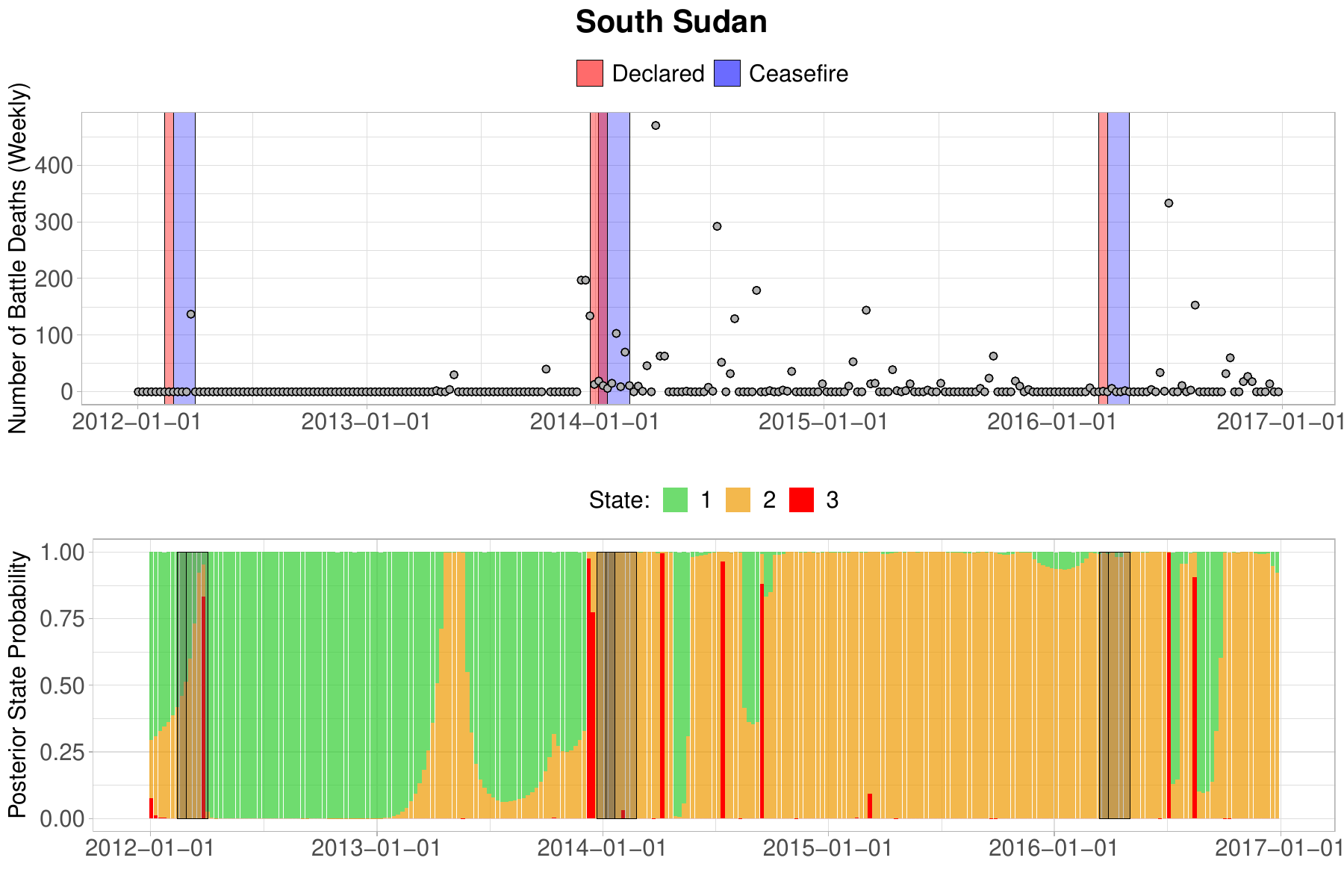}
\caption{\footnotesize The top panel displays the observed data for South Sudan, and the bottom panel displays the estimated posterior probability of each state for each week.}
\label{fig:PosteriorStateSouthSudan}
\end{figure}

\begin{figure}[H]
\centering\singlespacing
\includegraphics[scale=.45]{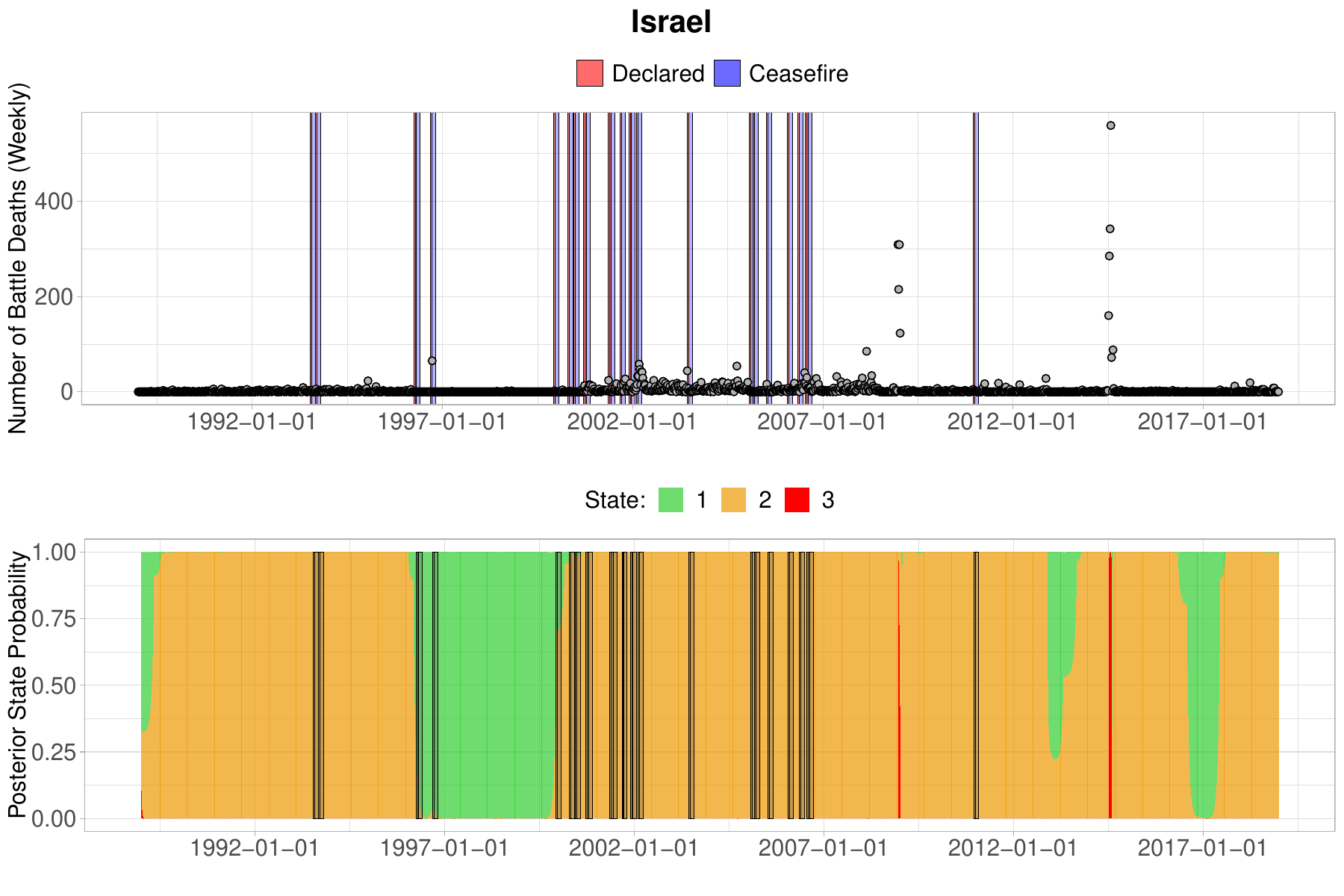}
\caption{\footnotesize The top panel displays the observed data for Israel, and the bottom panel displays the estimated posterior probability of each state for each week.}
\label{fig:PosteriorStateIsrael}
\end{figure}

For our analysis of the real data, we focus on three inferential aspects of the fitted model.  First, the posterior mean estimates of all 70 HMM parameters described in Section \ref{methods} are summarized in Table \ref{posterior_means}.  The MCMC trace plots and histograms of the posterior samples are provided in the Supplementary Material \citep{williams2024}.  The posterior means presented in Table \ref{posterior_means} quantify the effect of the various parameters/features in the model, but they are not dynamic in the sense that they represent the model at a weekly resolution whereas the HMM is fitted to the data as a system that evolves over many years of accumulated weeks.  For this purpose, second, we present probability evolution curves for a small selection of countries in Figures \ref{fig:transSouthSudan} and \ref{fig:transIsrael}.  For these figures, the transition probabilities are computed and plotted over the same time periods observed in the training data, using the covariate values associated with each country/week.  Furthermore, these probability evolution plots ignore the HMM response data (i.e., the counts of conflict deaths each week) to provide inference purely on the effects of the state transition probability covariates.  In particular, they demonstrate the role that ceasefires have played in the de-escalation of violence for the countries in our data set.  Third, we use the posterior mean estimates from Table \ref{posterior_means} to sample state sequences via the state space sampler discussed in Section \ref{sec:implementation}. These are displayed, for the same small selection of countries, in Figures \ref{fig:PosteriorStateSouthSudan} and \ref{fig:PosteriorStateIsrael}. The probability evolution and state space sampler plots for all 170 countries included in our data set are available with our Supplementary Material \citep{williams2024}.

\begin{table}[H]
\setlength{\tabcolsep}{1.2mm}
\begin{tabular}{p{0.1\linewidth} rrrrrrrr}
trans. rates & baseline & pre-ceasefire & ceasefire & v2x & v2x$^2$ & v2x$^3$ & GDP & pop \\
\hline 
$\zeta_{3}'$ ($2\to1$) & $\bf -4.714^{\star}$ & $-0.322$ & $\bf 1.243^{\star}$ & $\bf -0.938^{\star}$ & $\bf 1.748^{\star}$ & $\bf -0.899^{\star}$ & $\bf -0.554^{\star}$ & $\bf -0.588^{\star}$ \\
$\zeta_{4}'$ ($2\to3$) & $\bf -5.965^{\star}$ & $\bf 1.693^{\star}$ & $\bf 0.524^{\star}$ & $-0.375$ & $\bf -0.572^{\star}$ & $0.421$ & $\bf -0.486^{\star}$ & $\bf -0.516^{\star}$ \\
$\zeta_{5}'$ ($3\to1$) & $\bf -0.986^{\star}$ & $\bf -1.550^{\star}$ & $0.367$ & $\bf -1.617^{\star}$ & $0.153$ & $0.317$ & $\bf -0.554^{\star}$ & $-0.005$ \\
$\zeta_{6}'$ ($3\to2$) & $\bf 0.993^{\star}$ & $-0.289$ & $0.161$ & $\bf -0.734^{\star}$ & $0.134$ & $0.245$ & $0.143$ & $0.305$ \\
\\
AR coef. & baseline & pre-ceasefire & ceasefire & v2x & v2x$^2$ & v2x$^3$ & GDP & pop \\
\hline
$\beta_{1}'$ (state 2) & $\bf -4.228^{\star}$ & $0.078$ & $\bf -0.099^{\star}$ & $\bf 1.784^{\star}$ & $\bf -3.945^{\star}$ & $\bf 2.389^{\star}$ & $\bf 0.238^{\star}$ & $\bf 0.337^{\star}$ \\
$\beta_{2}'$ (state 3) & $\bf -3.849^{\star}$ & $\bf 0.660^{\star}$ & $0.061$ & $\bf -0.986^{\star}$ & $\bf -0.851^{\star}$ & $0.392$ & $\bf 0.724^{\star}$ & $\bf 0.948^{\star}$ \\
\\
other & $a_{1}$ & $a_{2}$ & $a_{3}$ & $c$ & $\pi_{2}$ & $\pi_{3}$ \\
\hline
& $\bf 0.0004^{\star}$ & $\bf 0.0911^{\star}$ & $\bf 5.8714^{\star}$ & $\bf 0.0246^{\star}$ & $\bf 0.0279^{\star}$ & $\bf 0.0140^{\star}$ \\
\end{tabular}
\caption{\footnotesize Posterior means of the HMM parameters. Note that parameter values reflect covariate values for lag v2x polyarchy (linear, quadractic, and cubic), lag and log GDP per capita, and lag and log population.  These variables have all been centered and scaled to have unit standard deviation. Boldface$^{\star}$ indicates that the .95 credible region, formed by excluding posterior samples in the upper and lower .025 tails, excludes the value 0. Note that we omit the state $1\to2$ transitions from this table because the inferential focus of our application to conflict research is restricted to the other transitions. See the Supplementary Material \citep{williams2024} for the MCMC trace plots and histograms of the posterior samples.}\label{posterior_means}
\end{table}

The parameter estimates in Table \ref{posterior_means} suggest a variety of interesting findings. First, we note that the state 2 (stable violence) to state 1 (non-violent) transition probability increases by about a factor of 3.5 when a ceasefire is in effect, taking all other covariates at mean value.  Note that while there is also found to be a statistically significant, positive coefficient for the ceasefire indicator variable associated with the $2\to3$ transition, the larger magnitude of the $2\to1$ coefficient indicates that $2\to1$ transitions will occur with higher probability than $2\to3$ transitions.  Nonetheless, the finding of positive coefficients for both the $2\to1$ and $2\to3$ transitions is evidence that ceasefires are directly associated with {\em some} change in the underlying conflict dynamics, most likely a cessation of violence.  Over time, the factor of 3.5 effect of ceasefires is visually displayed in the top panels of Figures \ref{fig:transSouthSudan} and \ref{fig:transIsrael}, within the vertical bars that indicate ceasefires are in effect. Note that it is also observed in the figures that this effect carries momentum for diminishing state 2 or 3 transition probabilities even after the ceasefire period.  Second, observe the statistically significant, positive pre-ceasefire indicator variable coefficient appearing for transition $2\to3$, as well as the statistically significant, negative pre-ceasefire indicator variable coefficient for the $3\to1$ transition.  Such findings suggest heightened or sustained levels of violence in the weeks associated with the pre-ceasefire period.  This is a major finding of our analysis.  It highlights the lag time between when a ceasefire is negotiated and when it actually begins, and that negotiating and preparing for a ceasefire is associated with an immediate short-term escalation in violence (which in turn is likely to also increase the likelihood of a ceasefire). 

Parties have incentives to fight harder to gain the strongest relative position prior to the ceasefire suspending the violence. Escalated violence also increases the incentives for a ceasefire. Once a ceasefire enters into effect, we find that conflict dynamics tend to transition from a violent to a non-violent state in the weeks that follow the ceasefire.  An explanation is that the benefits accrued from a ceasefire, whether peaceful or military/strategic, require some immediate shift in violence dynamics.  Finally, we find evidence for three underlying states of conflict, which we describe as `non-violent', `stable violence', and `intensified violence'.  We illustrate the utility of the constructed HMM for both inferential purposes and as a tool for predicting intensity and violence in conflict. 

Furthermore, the estimated mean AR coefficient, $\rho_{i,k} / c$ (recall equation (\ref{exp_value})), increases from 0.8659 in state 3 to 1.6753, taking all other covariates at mean value, for all pre-ceasefire weeks.  However, in either case, this coefficient will be explosive (i.e., the AR process is not stationary) for countries with larger GDP per capita and/or larger population, as demonstrated by the significant coefficient estimates 0.724 and 0.948, respectively. The explosive value of this coefficient is consistent with our interpretation of state 3 as `intensified' violence. Conversely, the `stable' violence interpretation for state 2 comes from the fact that it has an AR coefficient, taking all other covariates at mean value, estimated to be less than 1, which describes a weakly stationary process. Such processes revert to a stationary mean, and it is in this sense that the `stable' violence state describes both non-escalating {\em and} de-escalating violence.

One final important finding is the statistically significant effect for the third degree polynomial coefficient for the v2x polyarchy variable for the state 2 to 1 transition probability (with a leading negative coefficient), and for the state 2 AR coefficient.  Both democracies and autocracies are less likely than countries in the middle to enter into conflict, but when they do their trajectories differ dramatically.  Democracies are known to have less fatal conflicts than other regimes \citep{Lacina_2006JCR}, but also more durable conflicts \citep{Crisman-Cox_2022}.  Conflicts are more likely to erupt in the hybrid regimes between pure dictatorships and democracies \citep{hegre_2001}.  The significant effect of a third degree polynomial for the v2x polyarchy variable is evidence for the hypothesized non-monotonic nature of the association between measure of democracy and violent conflict dynamics.  Our results suggest that low-level conflicts are most likely to terminate for autocratic regimes, and most likely to escalate for hybrid regimes. Democracies are least likely to move in either direction, which is consistent with the literature.  For instance, consider the Middle East.  Israel, a democracy, has been involved in a series of armed conflicts with various non-state organizations. These conflicts usually remain active but do not escalate beyond a certain level because Israel puts a limit on its own use of force. Its most proximate neighbors, however, have not shown the same restraint.  Jordan used maximum force in September 1970 to expel PLO from Jordan, and were successful.  Four decades later, Syria tried the same strategy, and failed, with the Syrian Civil War as a consequence.

GDP per capita and population show very similar effects. It is important to keep in mind that conflicts tend to happen in poorer countries. Hence, the negative effect of GDP per capita on transitions and the positive effect on fatalities must be interpreted in the context of the sample of countries in the data set that had conflicts. The very poorest countries are more likely to have very intermittent conflicts, which causes the country to move often between state 1 and 2, and sometimes state 3. Richer countries are, once they find themselves in a state of conflict, more capable of persistently fighting these conflicts. Similarly, more populous countries are also more likely to maintain a strong military and are as such more likely to field a sustained campaign. This does not mean that the most fatal conflicts necessarily are in richer country, but the conflicts with one or two deaths per week, scattered over time tend to be in the poorest and smallest countries.

The fact that the data exhibit patterns of both stable and unstable models is a strong indication that conflicts endure distinct phases, and that intensified phases marked by explosive violence are only sustained for short periods of time.  Namely, observe that, taking all other covariates at mean value and in the absence of ceasefires or negotiations, the baseline transition probabilities from $3\to1$, $3\to2$, and $3\to3$ are 0.0916, 0.6628, and 0.2456, respectively; once in state 3 (intensified violence) it is about 0.75 probability of transitioning to another state in the next week. 

Our last remark on the estimates in Table \ref{posterior_means} is that the fitted expected number of battle deaths per week, in the absence of battle deaths in previous weeks, during state 1, state 2, and state 3 are 0.0163, 3.7033, and 238.6748, respectively.  Recall from the expected value of the negative-binomial distribution in (\ref{exp_value}) that these are computed as $a_{j} / c$, for $j \in \{1,2,3\}$.  This statistic serves as a simple, rudimentary description of how the HMM has fitted the mean behavior of each of the three states.

Finally, we note that (although not displayed in Table \ref{posterior_means}) the baseline probabilities for transition from state $1\to2$ and $1\to3$, in one week, are estimated to be 0.0006276 and $6\times10^{-7}$, respectively.  This is consistent with the fact that periods of violent conflict are rare events, and are not part of the natural progression of the state of affairs for the average country, (average in the sense of the covariates we consider).  However, we find that the transition probability from state $1\to2$ increases by a factor of about 52 for a country with a declared ceasefire, and by a factor of about 18 for a country with a ceasefire in effect, taking all other covariates at mean value.  Mostly, this reflects the fact that ceasefires occur in the midst of violent conflict, and so the likelihood of future violence is higher than if a country was in state 1 not having recently transitioned from a state of violent conflict.

\subsection{Addressing limitations of the data: model re-estimation using low and high estimates of weekly battle death counts}

UCDP states that the fatality estimates in the UCDP geo-referenced event data set used in this analysis are fraught with uncertainty \footnote{See https://www.pcr.uu.se/research/ucdp/methodology/ for a discussion.}. To quantify the uncertainty expressed in the source material, they publish three estimates: best, high, and low.  To clarify the UCDP geo-referenced event coding scheme \citep[][]{Hogbladh2023}, geo-referenced event events are defined based on a subjective judgement of one or more sources with at least one trustworthy claim that at least one person was killed.  The estimates of best, high, and low reflect the variance reported in the number of casualties from sources deemed trustworthy \citep[][]{Hogbladh2023}. UCDP has a conservative policy, by which they will report the lowest number supported by a source.
The variability between best, high, and low can arise either from a range reported by a source or different estimates given by different sources.  If a partially corresponding aggregate report contains a single fatality estimate, then the corresponding record for this event will be best = high = low.  There are generally four different types of uncertainty:
\begin{enumerate}
\item Some events are reported by several independent sources.  The September 11 terrorist attach in the United States, for instance, is reported by a large number of sources, which all report the same number of casualties (2,986). This is an example of a credible {\em and} precise number.
\item For some events, the best, high, and low estimates differ, sometimes by a lot.  There might be conflicting reports that claim either 2 dead or 25 dead. Both of these are precise, but they cannot both be credible.  Nonetheless, we have to choose between 2 or 25, not any random number in-between. 
\item There might be several sources that report, e.g., between 10 and 20 casualties. This is not precise, but can be credible. In this case any random number between 10 and 20 is as likely as any other.
\item There might be a single report of an imprecise number, such as ``at least 50,000 persons are killed''. This is not very precise, and it is hard to adjudicate the credibility.  We cannot report any number other than 50,000 without making a number of additional assumptions, but soley reporting 50,000 creates a false impression of precision.
\end{enumerate}

We use the best estimates in our main analysis in the previous section, but we check the robustness of the fitted model using both the high and low estimates, here; see Tables \ref{posterior_means_high} and \ref{posterior_means_low}.  We additionally fit the model on another 100 data sets, where the number of fatalities each week is taken as a mixture drawn from all three categories to assess the robustness of the current model to the unknown variability across these categories; see Table \ref{posterior_means_mixture}.  The HMM fit to the best--high--low mixture is largely consistent with the model fit using the UCDP best estimates of fatalities, even more so than the parameters fit using exclusively the UCDP high or low estimates.

\begin{table}[H]
\setlength{\tabcolsep}{1.2mm}
\begin{tabular}{p{0.1\linewidth} rrrrrrrr}
trans. rates & baseline & pre-ceasefire & ceasefire & v2x & v2x$^2$ & v2x$^3$ & GDP & pop \\
\hline  
$\zeta_{3}'$ ($2\to1$) & $\bf -4.572^{\star}$ & $-0.354$ & $\bf 0.947^{\star}$ & $\bf -0.954^{\star}$ & $\bf 1.924^{\star}$ & $\bf -1.069^{\star}$ & $\bf -0.620^{\star}$ & $\bf -0.619^{\star}$ \\
$\zeta_{4}'$ ($2\to3$) & $\bf -6.141^{\star}$ & $\bf 1.458^{\star}$ & $\bf 1.195^{\star}$ & $\bf -0.894^{\star}$ & $-0.104$ & $0.115$ & $\bf -0.490^{\star}$ & $\bf -0.472^{\star}$ \\
$\zeta_{5}'$ ($3\to1$) & $\bf 2.427^{\star}$ & $\bf -1.239^{\star}$ & $\bf -0.641^{\star}$ & $0.205$ & $-0.618$ & $\bf 1.507^{\star}$ & $0.040$ & $-0.385$ \\
$\zeta_{6}'$ ($3\to2$) & $\bf 1.372^{\star}$ & $0.046$ & $0.201$ & $-0.143$ & $0.145$ & $-0.352$ & $\bf 0.501^{\star}$ & $\bf 0.501^{\star}$ \\
\\
AR coef. & baseline & pre-ceasefire & ceasefire & v2x & v2x$^2$ & v2x$^3$ & GDP & pop \\
\hline  
$\beta_{1}'$ (state 2) & $\bf -4.404^{\star}$ & $-0.008$ & $\bf -0.219^{\star}$ & $\bf 2.133^{\star}$ & $\bf -4.543^{\star}$ & $\bf 2.732^{\star}$ & $\bf 0.239^{\star}$ & $\bf 0.321^{\star}$ \\
$\beta_{2}'$ (state 3) & $\bf -2.102^{\star}$ & $\bf -0.753^{\star}$ & $\bf -0.421^{\star}$ & $\bf -1.004^{\star}$ & $0.778$ & $0.543$ & $\bf 0.615^{\star}$ & $\bf 0.655^{\star}$ \\
\\
other & $a_{1}$ & $a_{2}$ & $a_{3}$ & $c$ & $\pi_{2}$ & $\pi_{3}$ \\
\hline
& $\bf 0.0004^{\star}$ & $\bf 0.0937^{\star}$ & $\bf 5.7008^{\star}$ & $\bf 0.0201^{\star}$ & $\bf 0.0396^{\star}$ & $\bf 0.0208^{\star}$ \\
\end{tabular}
\caption{Posterior means of the HMM parameters {\bf using high counts of battle deaths}. Compare with Table \ref{posterior_means}.}\label{posterior_means_high}
\end{table}

\begin{table}[H]
\setlength{\tabcolsep}{1.2mm}
\begin{tabular}{p{0.1\linewidth} rrrrrrrr}
trans. rates & baseline & pre-ceasefire & ceasefire & v2x & v2x$^2$ & v2x$^3$ & GDP & pop \\
\hline  
$\zeta_{3}'$ ($2\to1$) & $\bf -4.773^{\star}$ & $0.079$ & $\bf 1.228^{\star}$ & $\bf -1.328^{\star}$ & $\bf 2.618^{\star}$ & $\bf -1.506^{\star}$ & $\bf -0.508^{\star}$ & $\bf -0.560^{\star}$ \\
$\zeta_{4}'$ ($2\to3$) & $\bf -6.323^{\star}$ & $\bf 1.078^{\star}$ & $\bf 1.169^{\star}$ & $\bf -1.030^{\star}$ & $0.440$ & $-0.111$ & $\bf -0.402^{\star}$ & $\bf -0.384^{\star}$ \\
$\zeta_{5}'$ ($3\to1$) & $\bf 3.289^{\star}$ & $\bf -1.244^{\star}$ & $0.742$ & $\bf -0.807^{\star}$ & $\bf 1.062^{\star}$ & $0.357$ & $0.129$ & $\bf -1.519^{\star}$ \\
$\zeta_{6}'$ ($3\to2$) & $\bf 1.898^{\star}$ & $-0.332$ & $\bf 1.126^{\star}$ & $-0.004$ & $0.091$ & $\bf -1.035^{\star}$ & $0.776$ & $-0.204$ \\
\\
AR coef. & baseline & pre-ceasefire & ceasefire & v2x & v2x$^2$ & v2x$^3$ & GDP & pop \\
\hline 
$\beta_{1}'$ (state 2) & $\bf -4.160^{\star}$ & $\bf 0.116^{\star}$ & $\bf -0.102^{\star}$ & $\bf 2.232^{\star}$ & $\bf -5.035^{\star}$ & $\bf 3.101^{\star}$ & $\bf 0.223^{\star}$ & $\bf 0.317^{\star}$ \\
$\beta_{2}'$ (state 3) & $\bf -2.731^{\star}$ & $\bf 1.675^{\star}$ & $\bf -1.198^{\star}$ & $\bf -1.072^{\star}$ & $-0.235$ & $\bf 1.069^{\star}$ & $\bf 0.673^{\star}$ & $\bf 1.001^{\star}$ \\
\\
other & $a_{1}$ & $a_{2}$ & $a_{3}$ & $c$ & $\pi_{2}$ & $\pi_{3}$ \\
\hline
& $\bf 0.0005^{\star}$ & $\bf 0.0891^{\star}$ & $\bf 6.4379^{\star}$ & $\bf 0.0252^{\star}$ & $\bf 0.0600^{\star}$ & $\bf 0.0007^{\star}$ \\
\end{tabular}
\caption{Posterior means of the HMM parameters {\bf using low counts of battle deaths}. Compare with Table \ref{posterior_means}.}\label{posterior_means_low}
\end{table}

As pointed out above, the interpretation of isolated coefficients is difficult. The intended effect of a ceasefire can both be to reduce the fatalities in a given state of conflict and to increase the transition probability to a less intensified state of conflict.  The AR coefficients for ceasefire are strongly negative in Tables \ref{posterior_means_high}, \ref{posterior_means_low}, and \ref{posterior_means_mixture} for state 2, and as such provide stronger support than the main results.

The finding that ceasefires are preceded by an increased intensity does not find similarly robust support. Using the UCDP low estimate counts, the pre-ceasefire period is associated with a strong increase in the number of fatalities, whereas the high estimate counts result in the opposite association.  This finding is hard to explain as anything other than a reflection of the uncertainties in the data.  The model fit on both the high and low estimate counts as well as the best--high--low mixture, however, finds support for an increased transition probability from state 2 to state 3 prior to ceasefires, as well as a reduced transition probability from state 3 to state 1, all consistent with the findings in Table \ref{posterior_means}.

The UCDP high estimates have associated baseline transition rates consistent with less duration in state 3 than those of the best estimates, and while the estimate for $a_{3}$ based on the high estimates is similar to that based on the best estimate, the baseline AR coefficient for state 3 is markedly higher based on the UCDP high estimates.  This reflects UCDP's tendency towards conservative estimates.  Alternatively, using the UCDP low estimate counts compared to the best, the estimates for $a_{3}$ and the baseline AR coefficient for state 3 are both higher, but the low estimates have associated baseline transition rates consistent with even less duration in state 3.  This suggests more pronounced differences between the distributions of battle death counts for states 2 and 3, associated with the UCDP low estimates (i.e., battle death counts that are more flat over time with abrupt but short-lived spikes).

\begin{table}[H]
\setlength{\tabcolsep}{1.2mm}
\begin{tabular}{p{0.1\linewidth} rrrrrrrr}
trans. rates & baseline & pre-ceasefire & ceasefire & v2x & v2x$^2$ & v2x$^3$ & GDP & pop \\
\hline  
$\zeta_{3}'$ ($2\to1$) & -4.73(.23) & -.37(.80) & 1.08(.33) & -1.02(.45) & 1.78(.66) & -.90(.58) & -.55(.13) & -.59(.15) \\
$\zeta_{4}'$ ($2\to3$) & -5.93(.45) & 1.60(.45) & .50(.51) & -.45(.57) & -.66(.76) & .52(.56) & -.44(.19) & -.59(.21) \\
$\zeta_{5}'$ ($3\to1$) & .30(.63) & -1.42(.75) & .13(.77) & -1.33(.56) & .44(.63) & .84(.76) & -.52(.53) & -.30(.59) \\
$\zeta_{6}'$ ($3\to2$) & .58(.45) & -.28(.76) & .13(.74) & -.68(.58) & .12(.80) & -.18(.69) & .16(.39) & .43(.47) \\
\\
AR coef. & baseline & pre-ceasefire & ceasefire & v2x & v2x$^2$ & v2x$^3$ & GDP & pop \\
\hline 
$\beta_{1}'$ (state 2) & -4.31(.05) & .07(.05) & -.13(.05) & 1.93(.26) & -4.13(.63) & 2.48(.41) & .25(.02) & .34(.02) \\
$\beta_{2}'$ (state 3) & -3.83(.48) & .64(.85) & -.09(.59) & -.96(.56) & -.60(.78) & .46(.70) & .66(.25) & 1.08(.43) \\
\\
other & $a_{1}$ & $a_{2}$ & $a_{3}$ & $c$ & $\pi_{2}$ & $\pi_{3}$ \\
\hline
& .00(.00) & .09(.00) & 6.18(.32) & .02(.00) & .04(.03) & .04(.04) \\
\end{tabular}
\caption{Average of posterior means of each HMM parameter over 100 data sets, each data set constructed by uniformly sampling over the best, high, or low count of battle deaths, when variation exists.  Standard deviations of the posterior means are given in parentheses.  Compare with Table \ref{posterior_means}.}\label{posterior_means_mixture}
\end{table}

\section{Concluding remarks}\label{conclusion}

There are many research directions we hope to investigate in furthering this work.  Namely, we hope to distinguish between ceasefires of different types and scope, separating those ceasefires that have markedly different characteristics (e.g., cessation of hostility arrangements vis-a-vis preliminary ceasefires). A complimentary research direction is to include a covariate to account for the duration of a given conflict (i.e., time in state 2 and/or state 3).  This would allow us to address questions relating to whether conflicts tend to have limited persistence.  Finally, spatial correlations in conflict data (e.g., due to geopolitics, etc.) are well established \citep{Gleditsch2007}.  In part, they stem from the fact that certain conflict-inducing factors are present in the same area.  Poverty, for instance, is geographically clustered.  But they also relate through conflict dynamics.  Violent conflict can spill over from one country to another, such as between Rwanda and the Democratic Republic of Congo.  This might be because ethnic groups cohabit both sides of a border, or it might be because a government intervenes in another country to prevent the organization of armed resistance, such as the Israeli intrusion of Lebanon.  In any case, spatial correlations are a feature of these data that we plan to address in a subsequent, more complex formulation of our HMM approach, in our future research.

Further open research problems include the following:
\begin{enumerate}
\item We investigate the patterns that civil conflicts share across countries.  However, if labels were on the dyad level (i.e., pairs of armed and opposing actors) rather than the country level, would the same patterns emerge?  Are there additional patterns that would emerge?  What about patterns on the resolution of conflict labeled data?
\item In our analysis, we are able to estimate the effect of ceasefires for static time periods immediately surrounding the agreement.  Further methods are needed to determine the length of persistence of a given ceasefire.  
\item What are the effects of peacekeeping efforts on the sustainability of ceasefires?  
\item What are the effects of seasonality on the sustainability of ceasefires?  Certain countries and regions simply cannot maintain violence during certain parts of the year (e.g., due to seasonal weather patterns, the necessity of farming, etc.).
\end{enumerate}

\begin{supplement}

\stitle{Code and data to reproduce the empirical results presented in the manuscript.}

\sdescription{The saved MCMC output for the real data analysis and the simulation study are provided, but they can of course be reproduced by re-running the provided script files and data.  The figures presented in the paper, along with the figures delegated to the Supplementary Material are contained in these files.  See workflow.sh for line-by-line Unix command line code for reproducing the results.  Instructions for how the code should be run and parallelized are also provided in the workflow.sh file.}

\end{supplement}

\bibliographystyle{imsart-nameyear} 
\bibliography{references}       

\end{document}